\documentclass[12pt]{article}
\usepackage[margin=1in]{geometry}
\usepackage{multirow}
\usepackage{amssymb}
\usepackage{latexsym}
\usepackage{amsmath}
\usepackage{url}
\usepackage{xcolor}
\usepackage{graphicx}
\usepackage[numbers,sort&compress]{natbib}
\usepackage{authblk}
\usepackage{lineno}
\usepackage{cleveref}

\begin{document}

\title{Near-Field Combustion-Noise Source Dynamics in a Reacting Supersonic Temporal Mixing Layer}

\author[1]{Sriram P. Kalathoor\thanks{Corresponding author: sriram2@gatech.edu}}
\author[1]{Joseph C. Oefelein}
\affil[1]{Daniel Guggenheim School of Aerospace Engineering, Georgia Institute of Technology, Atlanta, Georgia 30332, USA}

\date{}

\begin{abstract}
Compressibility and chemical reactions in reacting flows provide source mechanisms for pressure fluctuations whose signatures depend on the flow, source distribution, and acoustic environment. Bounded flows can sustain strong feedback and narrowband tones, whereas boundary-free flows more often exhibit broadband source activity distributed across frequency. Near-field combustion-noise source dynamics are examined in a supersonic reacting hydrogen-air temporal mixing layer using high-fidelity time-resolved direct numerical simulation data. Pressure, heat-release, and dilatation fields are used to identify how localized reacting structures, compressive disturbances, broadband spectral content, and burst-driven temporal organization interact. The results show weak pressure--heat-release coherence concentrated within selected low-frequency bands, with no collapse onto a single dominant mode. Combustion intermittency modulates the near-field pressure response within these bands and is accompanied by burst-driven amplitude modulation and transient trajectory sensitivity, while the overall dynamics remain broadband and bounded. A planar source-radiation-potential projection further shows moderate low-frequency angular bias associated with the heat-release source distribution. Information-theoretic measures indicate that pressure fluctuations share more statistical structure with heat release than with dilatation. The analysis characterizes source-side organization and near-field pressure response in the sampled DNS plane, providing a basis for subsequent observer-based or acoustic-analogy radiation calculations.
\end{abstract}

\maketitle

\section{Motivation and Prior Work}

Combustion noise arises from the interaction of unsteady heat release, compressibility, and turbulent flow dynamics and has long been recognized as an intrinsic feature of reacting flows. The theoretical foundations of combustion-noise analysis trace back to Lighthill’s acoustic analogy, which identified fluctuating flow quantities as volumetric acoustic source terms in an otherwise homogeneous wave equation \cite{Lighthill1952}. Subsequent extensions to reacting flows clarified that unsteady heat release introduces an effective monopole-like source, while momentum, entropy, and dilatation fluctuations contribute dipole- and quadrupole-type mechanisms \cite{Goldstein2003}.

Historically, much of the combustion-noise literature has focused on confined or semi-confined systems such as ducts, gas-turbine combustors, and rocket engines. In these configurations, acoustic reflections and geometric confinement promote strong feedback between pressure oscillations and heat release fluctuations, often resulting in thermoacoustic instabilities characterized by narrowband spectral peaks \cite{LieuwenYang2005}. The Rayleigh criterion provides the classical organizing principle for these systems, stating that acoustic oscillations grow when pressure and heat release fluctuations are positively phased \cite{Rayleigh1878}. Extensive experimental and numerical studies have demonstrated that such feedback can lock the system into self-sustained oscillations with well-defined frequencies \cite{Dowling1997,Candel2002,HuangYang2009,Lieuwen2003}.

More recent work has emphasized that even broadband turbulent forcing can selectively excite weakly stable acoustic modes in confined systems through stochastic resonance mechanisms \cite{MagriJuniper2014}. While this blurs the distinction between tonal and broadband behavior in marginally stable combustors, such effects rely fundamentally on the presence of a resonant acoustic path and therefore do not generalize to boundary-free flows.

In contrast to confined flows such as in combustors, open shear flows such as free jets, mixing layers, and wakes are associated with broadband source and radiation signatures, with spectral energy distributed continuously across frequency. Studies of non-reacting compressible jets and shear layers have established turbulent eddies and convecting structures as the primary drivers of broadband aerodynamic noise \cite{Tam1986,Freund2001,BogeyBailly2006}. At supersonic conditions, additional contributions arise from compressibility effects and Mach-wave radiation, while the spectral character remains predominantly broadband. \cite{Fortune2004} Accordingly, the emphasis shifts from instability eigenmodes to broadband source organization, source-term localization, and intermittency.

When chemical reactions are introduced, additional noise-generation pathways emerge through unsteady heat release. Experimental and numerical studies of reacting jets and open flames consistently report broadband combustion noise dominated by stochastic heat release fluctuations, punctuated by intermittent high-amplitude events \cite{Crighton1975,Lieuwen2012}. Measurements of turbulent premixed flames outside unstable operating regimes further show weak coherence and no dominant frequency selection, reinforcing the fundamentally broadband character of combustion noise in open configurations \cite{Schuller2009}.

To diagnose combustion-noise mechanisms in open reacting flows, many studies employ source-term formulations derived from linearized wave equations, in which the time derivative of the heat release rate provides a monopole-like contribution \cite{PoinsotLele1992,Swaminathan2011}. Dilatation- and entropy-related terms can also play an important role, particularly in compressible and supersonic regimes \cite{ChuKovasznay1958,Ihme2017}.

Although the Rayleigh index was originally formulated in the context of confined thermoacoustic systems, it has been increasingly applied in a local or statistical sense to turbulent reacting flows without strong acoustic feedback. Spatially resolved and conditionally averaged Rayleigh indices reveal that positive pressure--heat release correlation occurs intermittently and in localized regions, even in the absence of global instability \cite{Balachandran2005,Schuermans2023}. These findings suggest that, in open flows, the Rayleigh index should be interpreted as a measure of local pressure--source phasing and energy exchange. Related studies of noisy transitions and multifractal combustion-noise signatures further motivate statistical descriptions of pressure--heat-release coupling before the emergence of coherent instability \cite{Kalathoor2017,Nair2014}.

Because broadband combustion noise is inherently non-stationary, time-averaged spectral analysis alone is insufficient to capture its temporal variability. Time–frequency methods such as short-time Fourier transforms and wavelet analysis have therefore become common tools in combustion-noise studies, revealing transient spectral bursts associated with localized combustion events \cite{IllingworthWaughJuniper2013}. Complementary intermittency measures, including burst detection and waiting-time statistics, further demonstrate that combustion noise is dominated by sporadic, high-amplitude events rather than periodic oscillations \cite{Stankovic2013}. At low Mach numbers, near-field effects on pressure-heat-release coherence have recently been studied using theory \cite{Ha2025} and experiments \cite{Ha2026} in order to characterize acoustic compactness.

Higher-order spectral measures, such as bicoherence, indicate weak nonlinear phase locking in broadband reacting flows, in contrast to strongly unstable thermoacoustic systems \cite{KimPowers1979}. Advances in computational aeroacoustics have further enabled direct interrogation of broadband noise mechanisms using high-fidelity simulations. Reviews by Bodony and Lele emphasized that broadband noise generation is spatially distributed and cannot be reduced to a small number of coherent modes \cite{BodonyLele2012}. DNS studies of compressible and supersonic reacting flows similarly report intermittent coupling among pressure, heat release, and dilatation, yielding broadband acoustic signatures in the absence of geometric confinement. \cite{IhmePitsch2012}

Dynamical-systems and information-theoretic tools have also been applied to distinguish between low-dimensional oscillatory dynamics and high-dimensional, turbulence-driven behavior. Phase-space embeddings, recurrence plots, and Poincar\'e sections generally reveal diffuse, aperiodic dynamics in open reacting flows \cite{Eckmann1987,KantzSchreiber2004}. Information-theoretic measures such as mutual information and transfer entropy further show that coupling between heat release and pressure is weak, intermittent, and directional rather than locked into sustained feedback loops \cite{Schreiber2000,Runge2019}. Recent time--frequency studies confirm that weak spectral peaks in averaged spectra often arise from sporadic, short-lived events rather than persistent oscillatory dynamics \cite{WaughJuniper2014}.

Prior work establishes that combustion noise in boundary-free reacting flows is broadband, intermittent, and spatially distributed, with weak and transient coupling among pressure, heat release, and dilatation. However, most existing studies focus on spatially developing flows, lower-Mach-number regimes, or more restricted views of source-response coupling. The present work builds on this foundation by examining a supersonic reacting temporal mixing layer using high-fidelity DNS and a combined physical and statistical analysis. The temporal mixing layer is used as a canonical source-side configuration, so the analysis centers on near-field pressure/source organization in the sampled DNS plane, with observer-dependent radiation requiring a separate acoustic-analogy or observer-surface treatment. Together, the source reconstruction, phasing, radiation-potential, spectral, intermittency, and state-space analyses quantify how localized heat-release events organize the broadband near-field pressure response in a boundary-free configuration.

\section{Numerical Method and Problem Setup}

We denote density by $\rho$, velocity by $\mathbf{u}=(u,v,w)$, pressure by $p$, temperature by $T$, and total energy by $E$. Species mass fractions are $Y_k$ with reaction rates $\dot{\omega}_k$, the mixture gas constant is $R_{\mathrm{mix}}$, and the ratio of specific heats is $\gamma$. The mixture fraction is $Z$, and $\nabla\cdot\mathbf{u}$ denotes dilatation. Angle brackets denote averages, with explicit subscripts used when the averaging operation must be distinguished.

A standard finite volume framework is used to simulate the flow in this work, and RAPTOR, a massively parallel, fully compressible reacting DNS/LES solver, is used to perform the simulations. The solution procedure involves solving the fully coupled system of conservation equations for mass, momentum, energy, species, and thermodynamic state. Because the present work analyzes postprocessed pressure, heat-release, and dilatation fields from an established DNS configuration, we summarize only the elements needed for the acoustic-source interpretation and refer the reader to Oefelein~\cite{Oefelein2006-PAS} for a detailed description of the solver, its implementation, numerics, and capabilities.

The supersonic reacting hydrogen-air temporal mixing layer simulated as a DNS in this study follows \cite{Obrien2014}. The flow is allowed to evolve temporally, with a standard rescaling procedure that permits continuous temporal growth. The shear layer reaches sufficiently high temperatures, partly because of the hot oxidizer stream and partly because of shocks at supersonic conditions, to cause autoignition when the fuel and oxidizer streams mix. Heat release is computed using the species enthalpies $h_k$ and reaction rates $\dot{\omega}_k$. The present analysis uses mid-plane fields and full-plane averages over the sampled $(x,y)$ plane unless otherwise noted. Since the domain is a boundary-free temporal mixing layer, quantities labeled as acoustic response or impedance are treated as near-field pressure--velocity/source-response indicators for the sampled plane. Far-field radiation, duct modes, and boundary impedance require observer surfaces, radiation boundaries, or acoustic-analogy propagation that are outside the present DNS data product. Finally, the local speed of sound is constructed using the species composition, temperature, and ideal-gas assumption.

\section{Combustion Noise Characterization}
We use nondimensional variables based on averaged shear-layer thickness (based on vorticity) $\delta_0$ and the average streamwise convective velocity $U_c$,
\begin{equation}
  x^\star=\frac{x}{\delta_0},\quad t^\star=\frac{tU_c}{\delta_0},\quad
  p^\star=\frac{p-\langle p\rangle_{xy}}{\rho_0 U_c^2},\quad
  \dot{q}^\star=\frac{\dot{q}-\langle\dot{q}\rangle_{xy}}{\rho_0 U_c^3/\delta_0},
\end{equation}
where $U_c$ is the convective velocity (computed as the average absolute difference between the top and bottom streams), with $(\nabla\cdot\mathbf{u})^\star=(\delta_0/U_c)\nabla\cdot\mathbf{u}$ and $|\nabla Z|^{2\star}=\delta_0^2|\nabla Z|^2$. This normalization permits direct comparison across fields and ensures that time derivatives use the consistent operator $\partial_{t^\star}=(\delta_0/U_c)\partial_t$.

\subsection{Source Terms, Rayleigh Index, and Spatial Structure}
We analyze source-side indicators motivated by the linearized wave equation. The heat-release source proxy and Rayleigh index are, respectively
\begin{align}
  S^\star &= \frac{S}{\dot{q}_{\mathrm{ref}}U_c/\delta_0} = \dfrac{(\gamma-1)}{\dot{q}_{\mathrm{ref}}U_c/\delta_0}\frac{\partial \dot{q}}{\partial t};\label{eq:source}\\ \mathcal{R}^\star&=\frac{\mathcal{R}}{(\rho_0 U_c^2)(\rho_0 U_c^3/\delta_0)} = \frac{\langle p'\,\dot{q}' \rangle}{(\rho_0 U_c^2)(\rho_0 U_c^3/\delta_0)}\label{eq:rayleigh}
\end{align}
with conditional Rayleigh index variants based on high $|\nabla Z|^{2\star}$ and high $Y_{HO_2}$ to isolate flame-zone contributions (\cite{Kang2007}). We also compute dilatation forcing as $\partial_{t^\star}(\nabla\cdot\mathbf{u})^\star$ (\cite{Dowling1995}), and a pressure-fluctuation energy proxy as
\begin{equation}
  E_p=\frac{p'^2}{\rho c^2},\qquad E_p^\star=\frac{E_p}{\rho_0 U_c^2},
\end{equation}
as a compact broadband indicator of pressure-fluctuation intensity.

These quantities are used as source-side measures rather than as a closed acoustic analogy. In particular, $S^\star$ measures the local unsteadiness of heat release with the thermodynamic factor appearing in the linearized source form, while the Rayleigh index measures local pressure--heat-release phasing in the sampled plane. The interpretation therefore concerns where and when combustion, compressibility, and pressure fluctuations organize in the DNS field.

\begin{figure*}[htbp]
  \centering
  \includegraphics[width=0.34\textwidth]{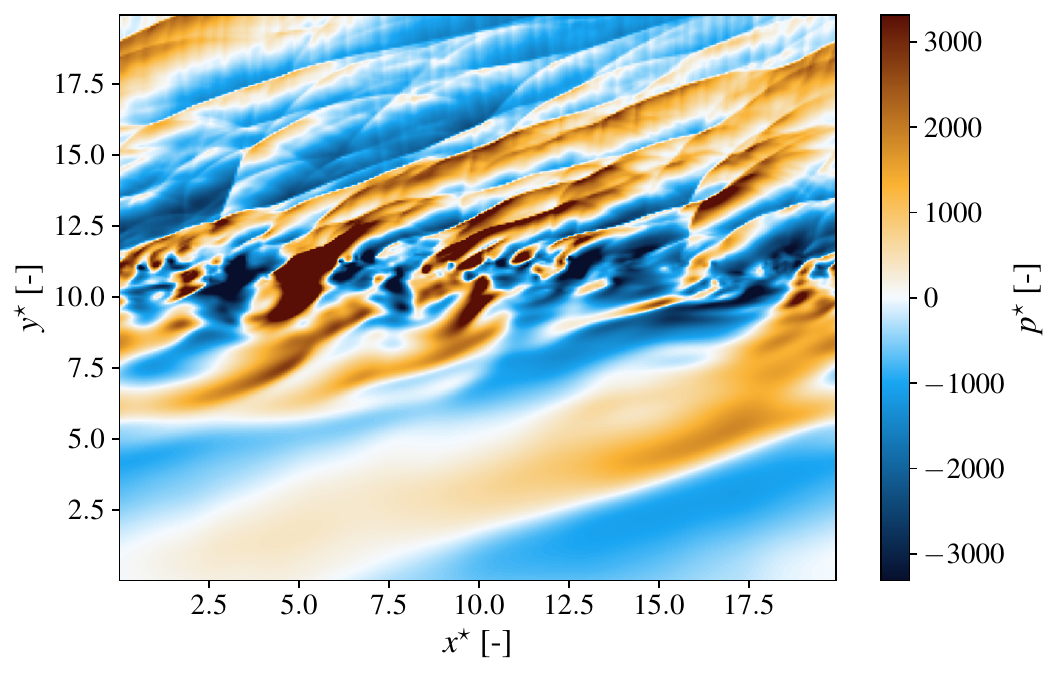}
  \includegraphics[width=0.34\textwidth]{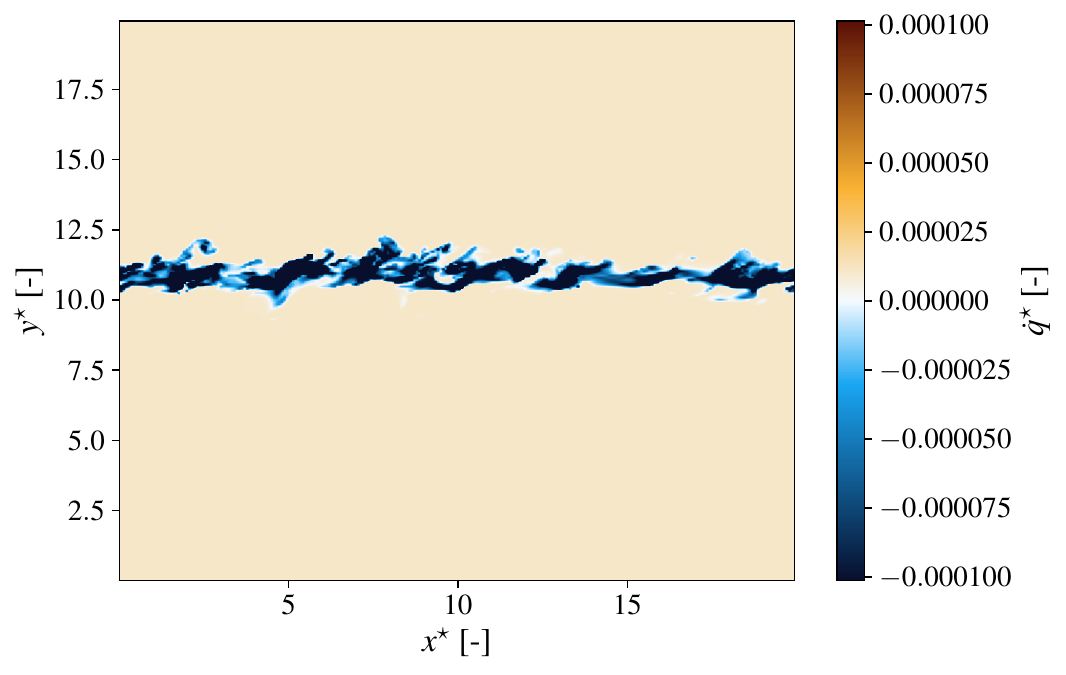}
  \\
  \includegraphics[width=0.34\textwidth]{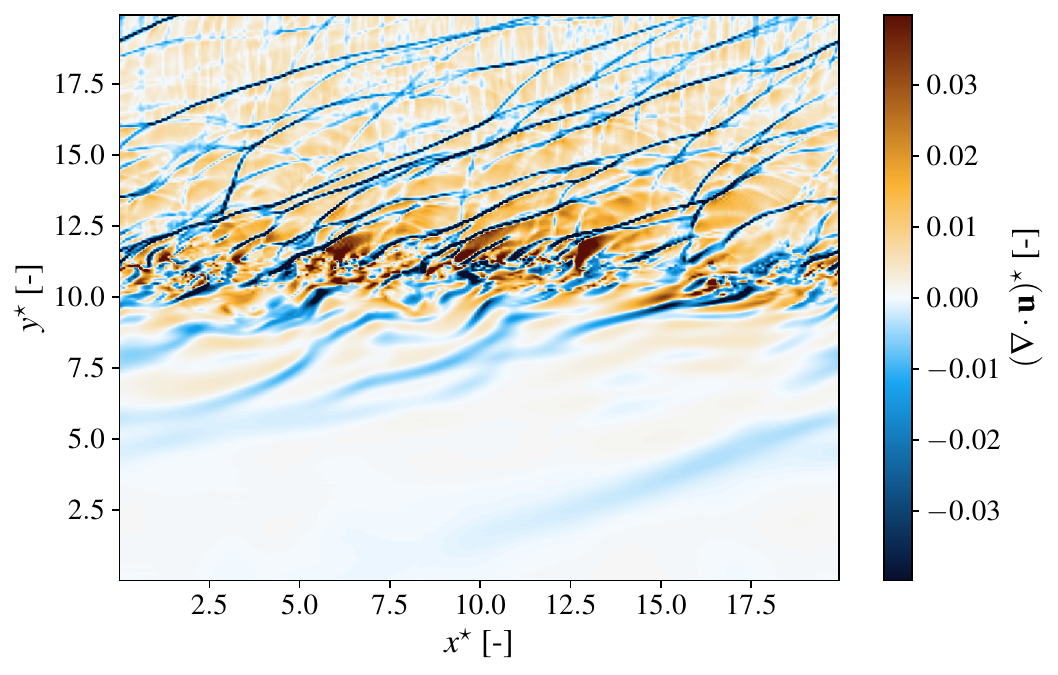}
  \includegraphics[width=0.34\textwidth]{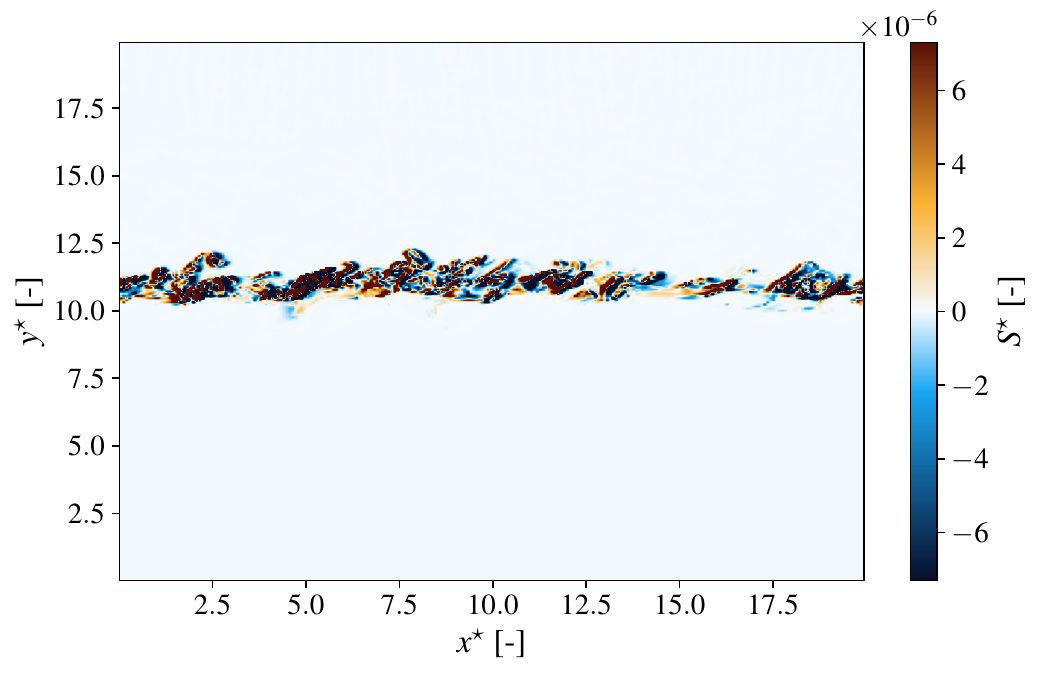}
  \\
  \includegraphics[width=0.34\textwidth]{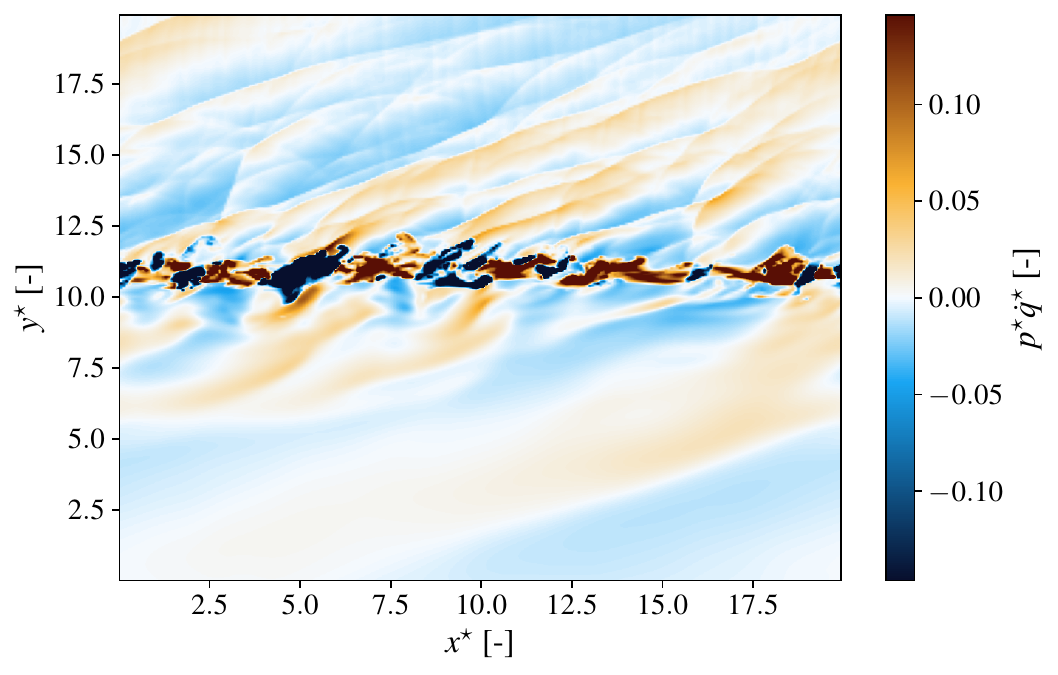}
  \includegraphics[width=0.34\textwidth]{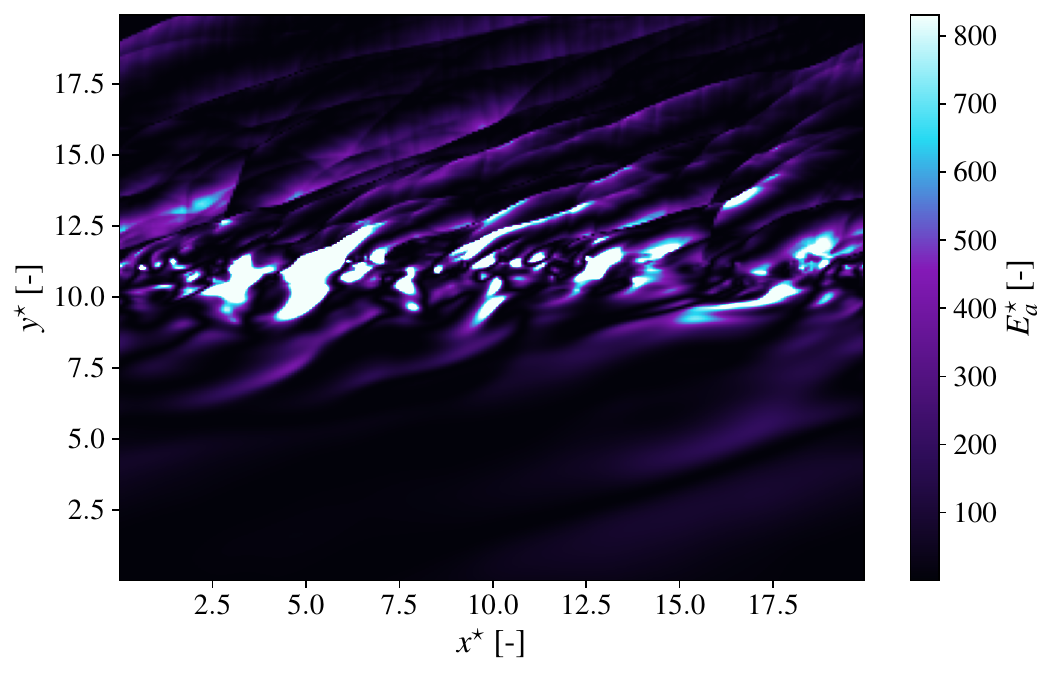}
  \caption{Mid-plane fields: first row, $p^\star$ and $\dot{q}^\star$; second row, $(\nabla\cdot\mathbf{u})^\star$ and heat-release source $S^\star$; third row, Rayleigh field $p^\star\dot{q}^\star$ and pressure-fluctuation energy proxy $E_p^\star$.}\label{fig:spatial_sources}
\end{figure*}

Mid-plane instantaneous visualizations of these quantities, shown in \Cref{fig:spatial_sources}, illustrate the spatial distribution of the primary source-side fields. As expected, heat release and its corresponding source proxies are confined to the vicinity of the reaction region, whereas dilatational and pressure fluctuations span much larger portions of the domain. As a result, the Rayleigh field is stronger in the reaction vicinity and weaker outside, while the corresponding pressure-fluctuation energy proxy is also elevated near the shear-layer core. In particular, the heat-release source shows limited spatial overlap with the broader shock-expansion pattern. This indicates that thermochemical activity is spatially localized within the reacting layer, whereas compressive and pressure signatures remain more distributed across the domain. However, the Rayleigh index, while larger in the chemically active regions, also exhibits both positive and negative spots, indicating spatially intermittent pressure--heat-release exchange rather than a uniformly phased acoustic response.

An important implication of the source-type decomposition is that it indicates whether reduced-order models should emphasize localized source regions or more distributed volumetric production. This distinction also affects how the near-field source organization should be interpreted, since compact and spatially distributed source patterns would enter any subsequent radiation calculation differently.

\begin{figure}[htbp]
  \centering
  \includegraphics[width=0.56\linewidth]{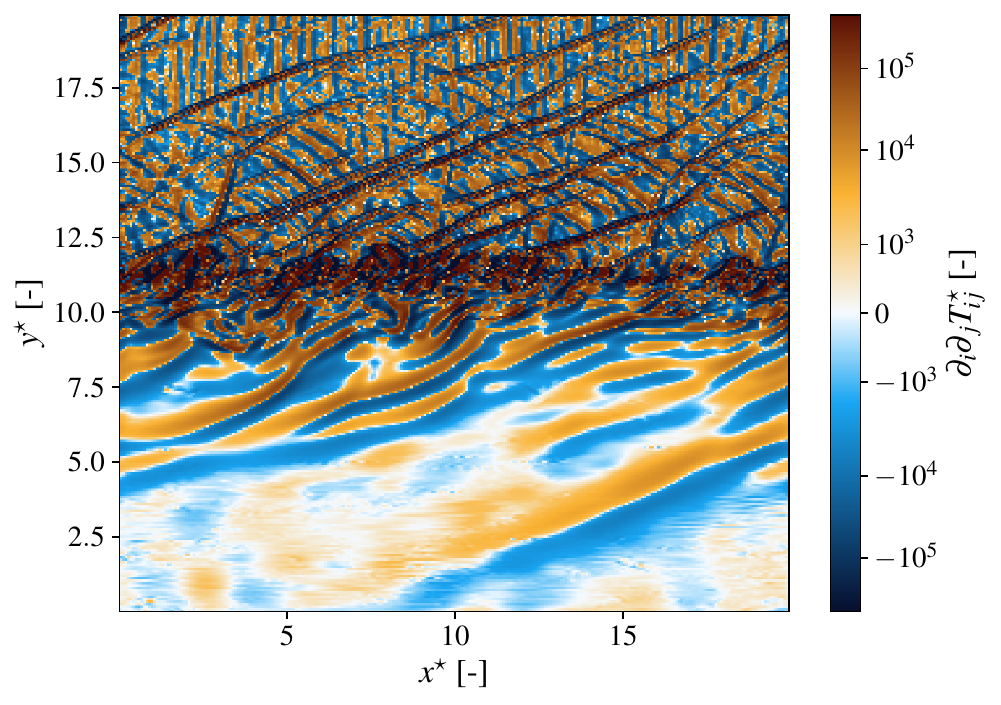}
  \\
  \includegraphics[width=0.56\linewidth]{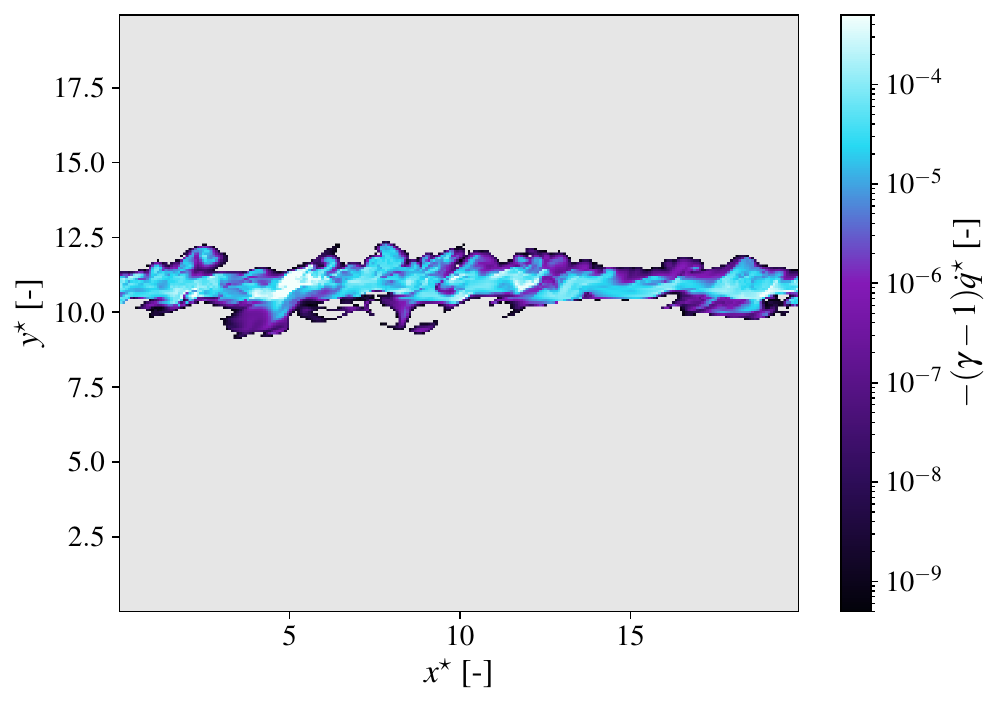}
  \\
  \includegraphics[width=0.56\linewidth]{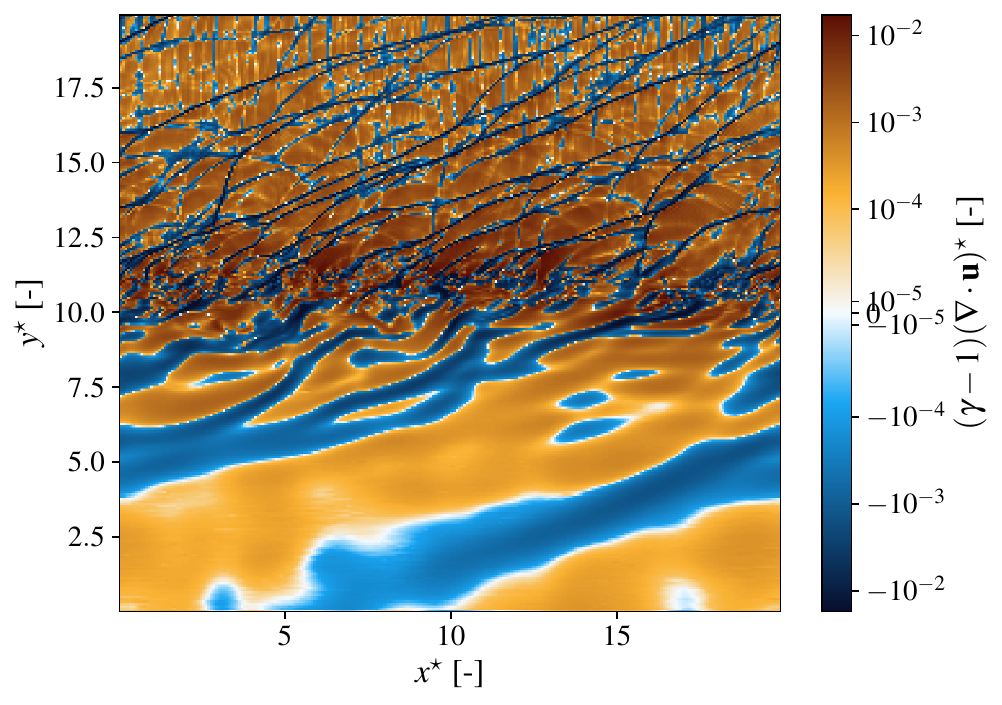}
  \caption{Source-type decomposition at a representative time instant: Lighthill-type quadrupole (top), heat-release monopole (middle), and dilatation source (bottom).}\label{fig:source_decomposition}
\end{figure}

\Cref{fig:source_decomposition} separates the source field into quadrupole, monopole, and dilatation-driven components. Compactness trends indicate that the strongest source-proxy regions remain spatially localized to within a small vicinity of the shear layer and the reaction zone. Under our non-dimensionalization, the Lighthill-type sources appear orders of magnitude stronger than the others. The Lighthill-type source is constructed as a double-divergence, whose derivatives act like a high-pass spatial filter in that they amplify small spatial scales and sharp gradients (such as shear layers, shocklets, steep strain regions), making the magnitudes large. Accordingly, the comparison is used to assess spatial support and source morphology; a radiation ranking would require a propagated acoustic field.

The spatial fields describe how localized events populate the shear layer and how their structure differs among variables. Pressure fluctuations are comparatively smoother than the scalar- and chemistry-driven quantities, reflecting the broader spatial support of compressive disturbances in the simulated domain. This is attributed to the supersonic freestream behavior, manifesting as large-scale shocks and expansions spanning much of the domain. In contrast, heat release and $|\nabla Z|^2$ respond directly to small-scale mixing and reaction activity, which yields finer spatial texture. The spatial coincidence of strong $p^\star$, $\dot{q}^\star$, and $(\nabla\cdot\mathbf{u})^\star$ events marks regions where combustion and compressibility interact most strongly, although it does not by itself establish a unique local driving mechanism.  We can characterize the locality of these effects using compactness measures.  The compactness length scales are computed from instantaneous mid-plane fields of $\dot{q}$ and $\nabla\cdot\mathbf{u}$. For each field $\phi(x,y)$, the spatial mean is first removed, giving $\phi'=\phi-\langle\phi\rangle_{x,y}$. The two-dimensional autocorrelation is then evaluated using the Fourier form
\begin{equation}
R_{\phi\phi}(\Delta x,\Delta y)
=
\mathcal{F}^{-1}\!\left[
\widehat{\phi'}\,\widehat{\phi'}^{\,\ast}
\right],
\end{equation}
and normalized so that $R_{\phi\phi}(0,0)=1$. One-dimensional cuts through the zero-lag center of the autocorrelation field are then extracted in the streamwise and cross-stream lag directions, $R_{\phi\phi}(\Delta x,0)$ and $R_{\phi\phi}(0,\Delta y)$. These cuts are not physical centerline profiles; they are lag-space measures of the spatial decorrelation of the full mid-plane field. The compactness lengths $L_x$ and $L_y$ are then defined as the first positive-lag locations where these cuts first fall below $e^{-1}$:
\begin{equation}
L_x = \min_{\Delta x>0}
\left\{
\Delta x : R_{\phi\phi}(\Delta x,0)<e^{-1}
\right\},
\qquad
L_y = \min_{\Delta y>0}
\left\{
\Delta y : R_{\phi\phi}(0,\Delta y)<e^{-1}
\right\}.
\end{equation}
The reported values are nondimensionalized by the reference length scale used in the simulation. Smaller values therefore indicate more compact source patches, whereas larger values indicate spatially broader coherent structure.

\begin{figure}[htbp]
    \centering
    \includegraphics[width=0.99\linewidth]{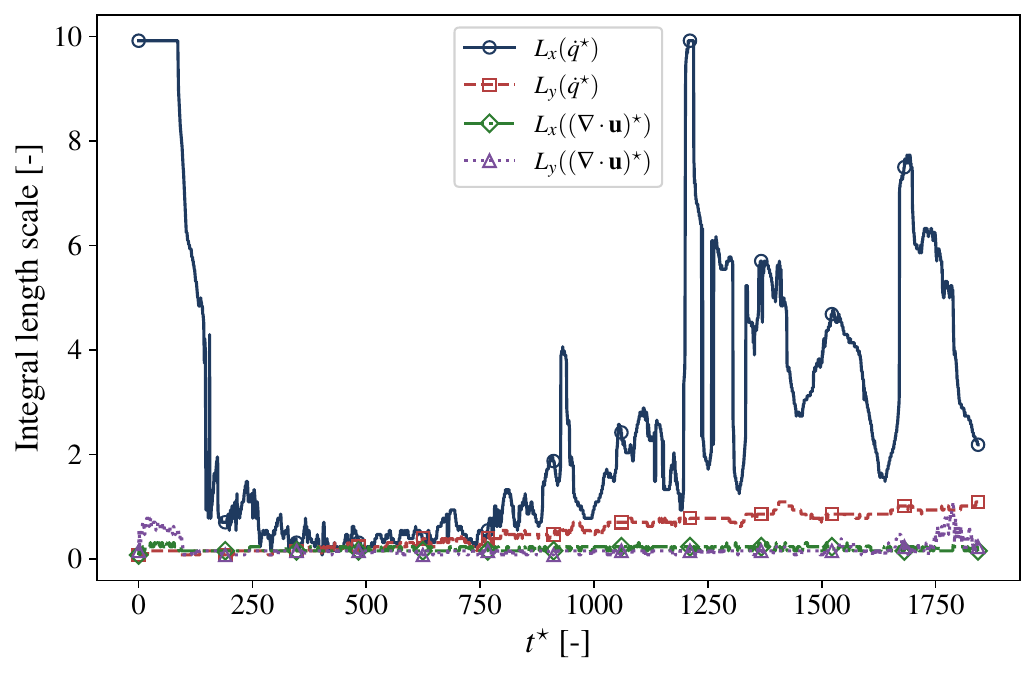}
    \caption{Compactness length scales ($L_x^\star$, $L_y^\star$) associated with heat-release fluctuations and dilatation}
    \label{fig:compactness}
\end{figure}

The compactness measures manifest as length scales that can be compared against the vorticity thickness. Length scales of order unity indicate that the strongest source-proxy patches remain confined to the shear-layer thickness rather than growing into a domain-scale coherent structure. \Cref{fig:compactness} shows that behavior, supporting a picture of localized, drift-aligned source activity rather than sustained resonance.

From a physical standpoint, these spatial fields also indicate how turbulence and chemistry couple. Regions of elevated $|\nabla Z|^2$ mark intense mixing, and when those regions coincide with strong $\dot{q}^\star$ they reveal zones of rapid heat release associated with local pressure fluctuations. The fact that these zones remain embedded in the shear layer suggests that the strongest source activity is tied to mixing-layer dynamics rather than to a large-scale acoustic feedback loop.

To connect the source morphology to an acoustic quantity, we compute a planar source-radiation-potential projection from the heat-release monopole proxy. Let $S_m=(\gamma-1)\dot{q}$ and let $\widehat{S_m'}(x,y;f^\star)$ denote the windowed temporal Fourier component of its fluctuation field over the available post-ignition planar record. For an observer direction $\theta$ in the sampled plane, the normalized source potential is based on
\begin{align}
  \mathcal{D}_m(\theta,f^\star)&=|\mathcal{A}_m(\theta,f^\star)|^2,\\
  \mathcal{A}_m(\theta,f^\star)&=
  2\pi f^\star
  \iint_{\Omega_{xy}}
  \widehat{S_m'}(x^\star,y^\star;f^\star)
  \exp\!\left[-i k^\star(f^\star)
  \left(x^\star\cos\theta+y^\star\sin\theta\right)\right]
  \mathrm{d}x^\star\mathrm{d}y^\star,
\end{align}
where $k^\star=2\pi f^\star M_c$, and $M_c$ is formed from the mean convective speed and a representative sound speed. This quantity measures the angular radiation potential of the sampled planar source distribution. It is interpreted at the source level, while absolute sound-pressure levels require a retarded-time acoustic analogy, three-dimensional observer surface, or radiation-boundary treatment.

\begin{figure*}[htbp]
    \centering
    \includegraphics[width=0.9\textwidth]{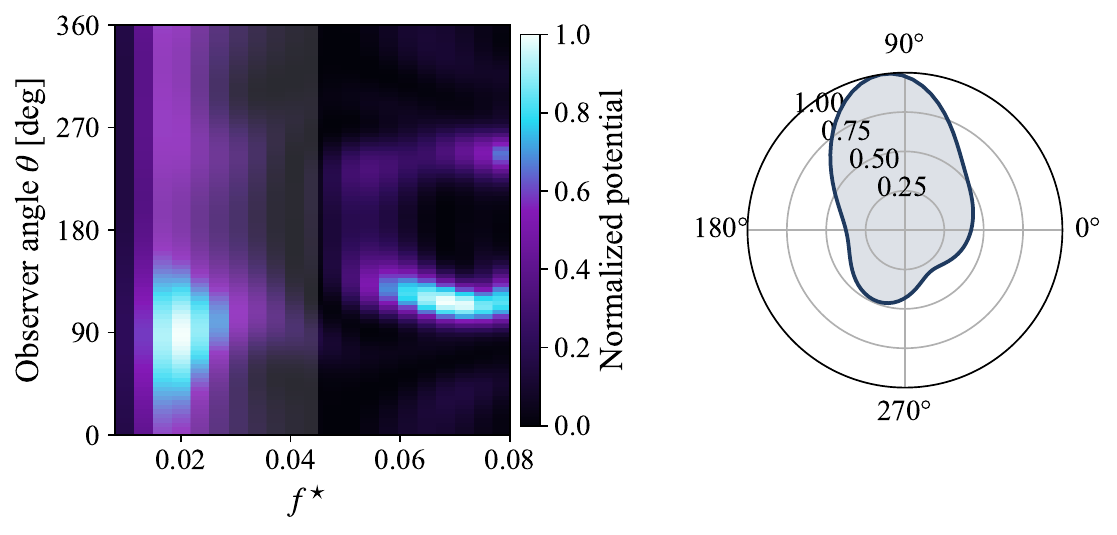}
    \caption{Planar heat-release source-radiation-potential proxy. The left panel shows normalized $\mathcal{D}_m(\theta,f^\star)$ from the windowed heat-release monopole source projected over observer angle in the sampled plane; the shaded frequency interval marks $0.015\le f^\star\le0.045$. The right panel shows the corresponding band-integrated angular distribution, normalized by its maximum.}
    \label{fig:radiation_potential}
\end{figure*}

The resulting angular distribution in \cref{fig:radiation_potential} shows that the heat-release source has a moderate low-frequency angular bias. Over the band $0.015\le f^\star\le0.045$, which contains the main pressure--heat-release coherence peaks, the band-integrated potential peaks near $\theta\simeq99^\circ$ and has a directivity index of approximately $2.8$ dB. The streamwise ratio $\mathcal{D}_m(0^\circ)/\mathcal{D}_m(180^\circ)\simeq1.1$ is close to symmetric, while the transverse ratio $\mathcal{D}_m(90^\circ)/\mathcal{D}_m(270^\circ)\simeq2.3$ indicates a stronger cross-stream bias. This supports the physical picture that the localized heat-release source patches are acoustically relevant and that their planar radiation potential remains broadband and moderately directional.


\subsection{Spectral and time--frequency analysis}

\begin{figure*}[htbp!]
  \centering
  \includegraphics[width=0.45\textwidth]{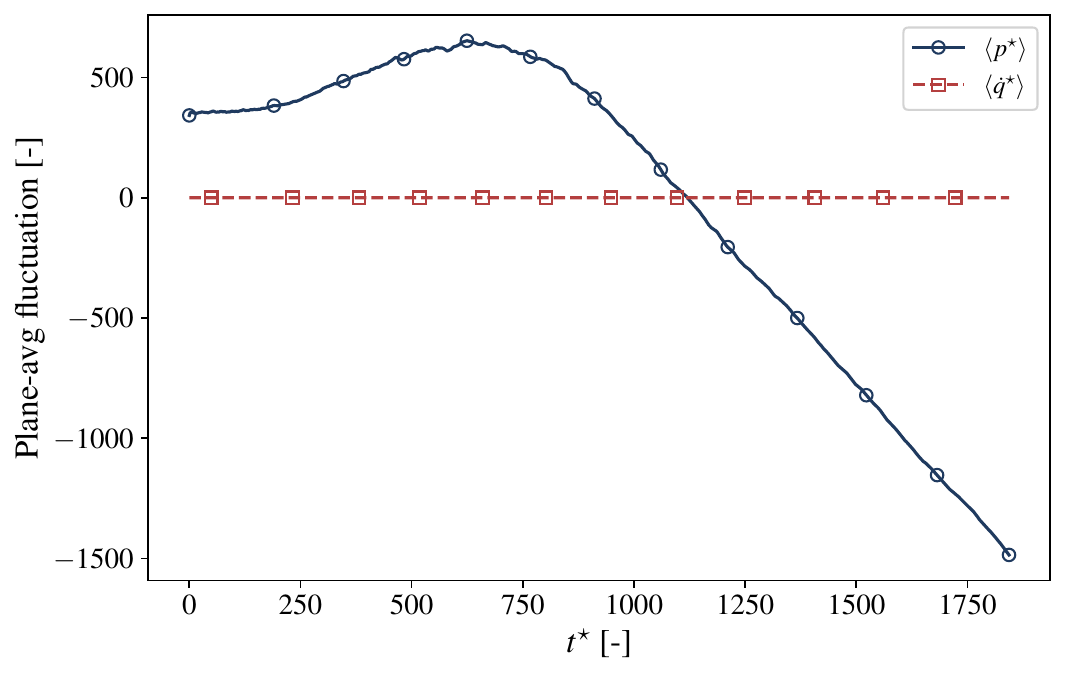}
  \includegraphics[width=0.45\textwidth]{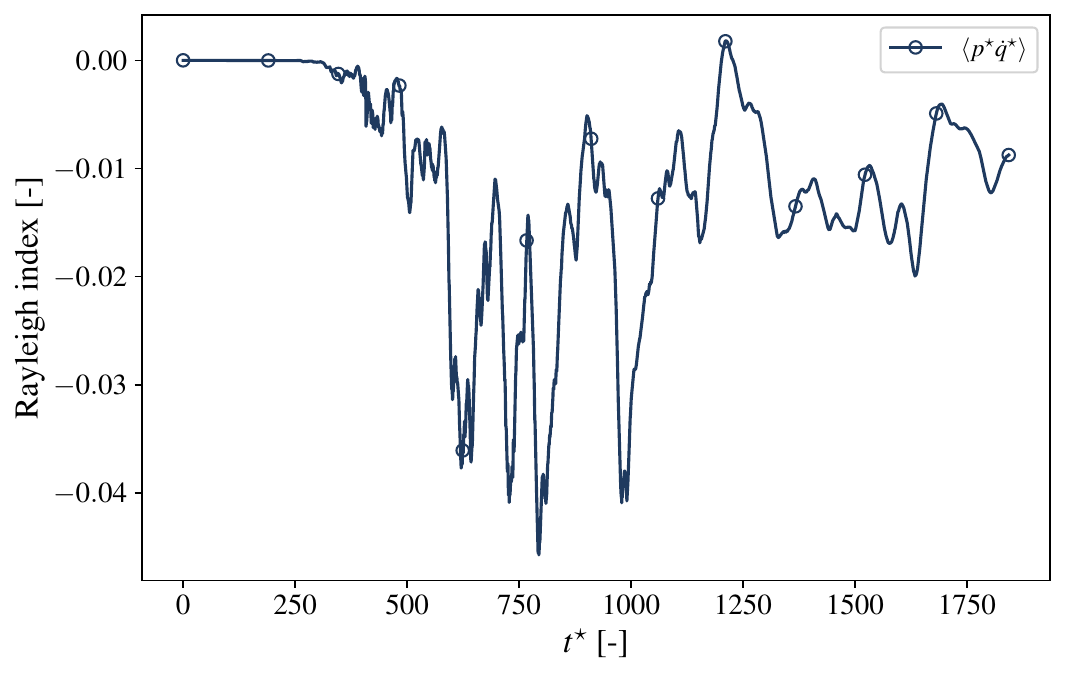}
  \\
  \includegraphics[width=0.45\textwidth]{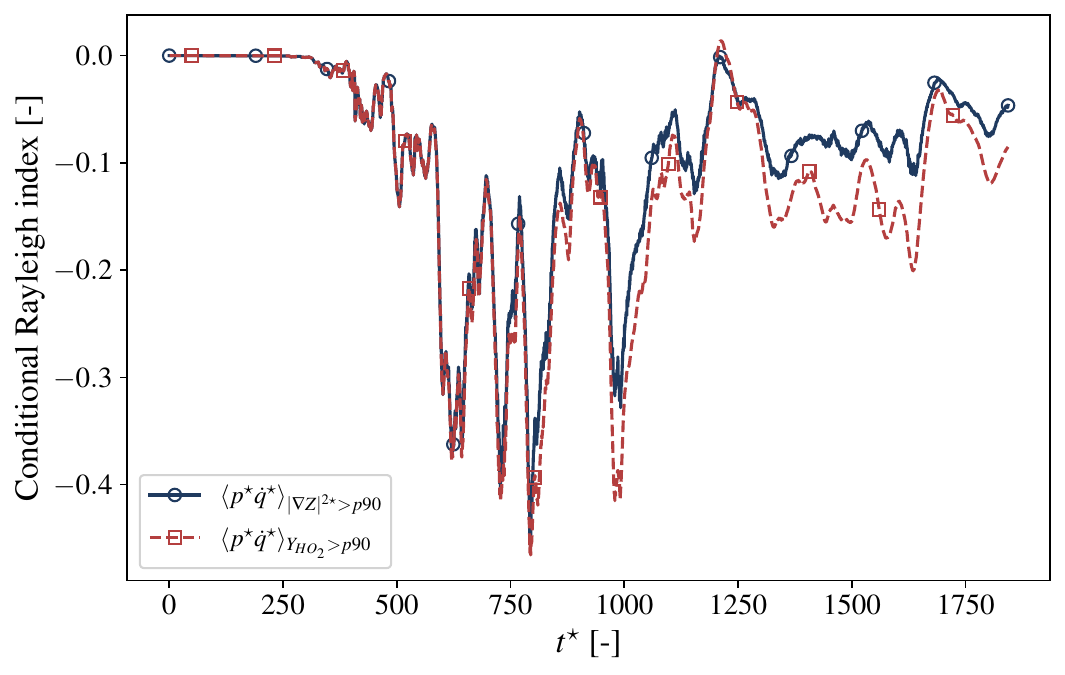}
  \includegraphics[width=0.45\textwidth]{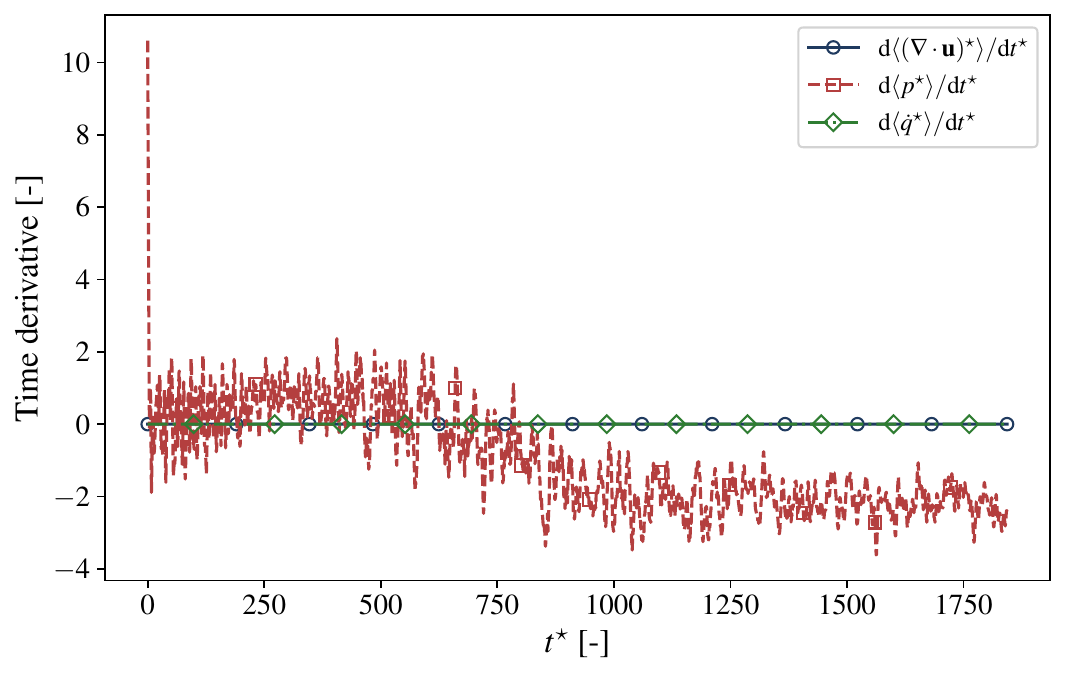}
  \caption{Plane-averaged time series: $p^\star$ and $\dot{q}^\star$, Rayleigh index, conditional Rayleigh index, and time derivatives.}\label{fig:timeseries}
\end{figure*}

Since the reacting shear layer evolves temporally, time-series quantification provides useful insight into the flow-reaction coupling and the near-field pressure response. Plane-averaged fluctuations and the Rayleigh index (both raw and conditional) shown in \cref{fig:timeseries} indicate weak pressure--heat release correlation with variable phase rather than a persistent phase-locked relationship. This is consistent with the absence of duct acoustics and with the broader shock-expansion pattern spanning the domain in contrast to the spatially localized reaction zones. The time-derivative signals support the same picture: plane-averaged dilatation and heat release remain comparatively weak and localized, while pressure is less localized and shows more transient temporal variability.

Because the plane-averaged quantities include the full sampled mid-plane, including freestream and weakly reacting regions, we additionally compare full-plane averages against reaction-zone-conditioned averages. The reaction-zone mask is defined independently at each time by the upper 10\% of $|\nabla Z|^2$, consistent with the scalar-gradient conditioning used in the Rayleigh-index analysis. This comparison separates the full-plane pressure/source response from the local behavior of the chemically active layer.

\begin{figure}[htbp]
    \centering
    \includegraphics[width=0.48\textwidth]{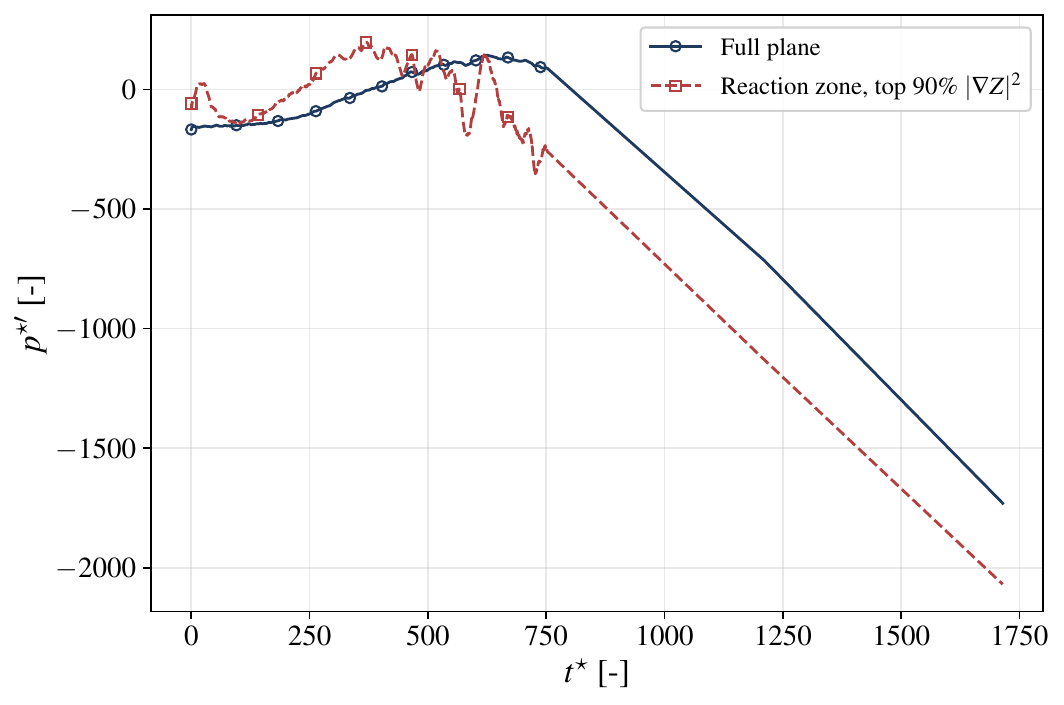}
    \caption{Comparison of full-plane and reaction-zone-conditioned pressure fluctuations. The reaction-zone mask is defined by the upper 10\% of $|\nabla Z|^2$ at each time.}
    \label{fig:plane_vs_rz_pressure}
\end{figure}

\begin{figure}[htbp]
    \centering
    \includegraphics[width=0.48\textwidth]{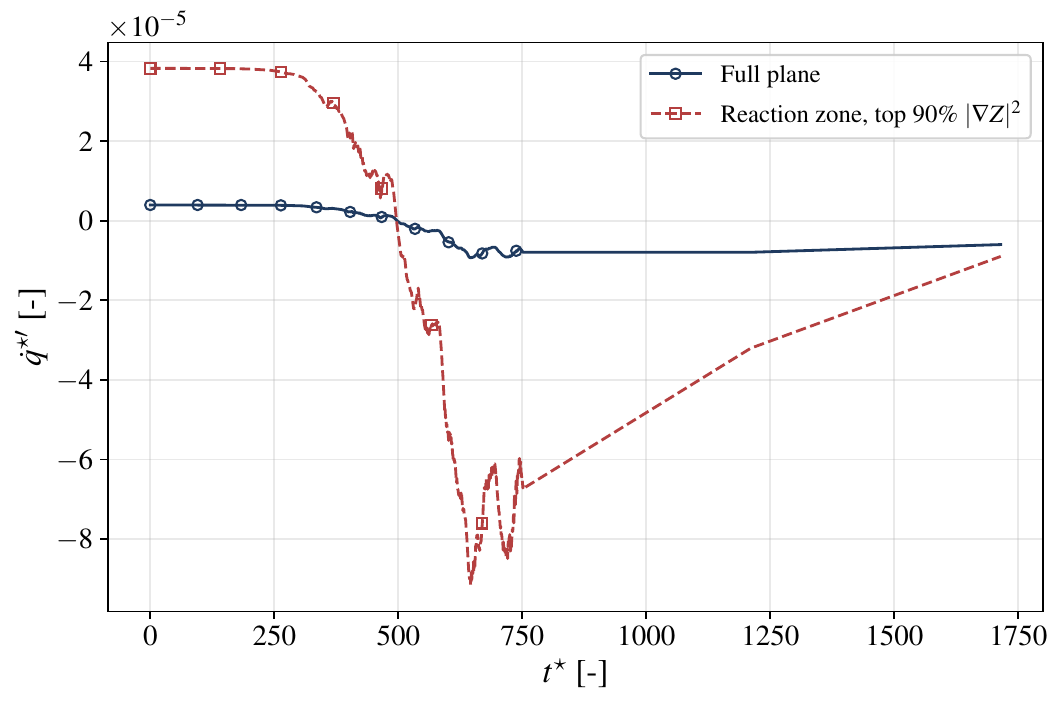}
    \caption{Comparison of full-plane and reaction-zone-conditioned heat-release fluctuations. Reaction-zone conditioning preserves the temporal structure of heat release while removing full-plane dilution by weakly reacting regions.}
    \label{fig:plane_vs_rz_qdot}
\end{figure}

The pressure comparison in \cref{fig:plane_vs_rz_pressure} shows that reaction-zone conditioning changes the pressure signal moderately in amplitude, with the conditioned pressure fluctuation RMS approximately 1.23 times the full-plane value. The full-plane and reaction-zone pressure traces are only weakly correlated, indicating that the full-plane pressure average contains substantial contributions from broader compressive and freestream regions outside the reaction zone. In contrast, \cref{fig:plane_vs_rz_qdot} shows nearly identical temporal structure between the full-plane and conditioned heat-release signals, but with a much larger conditioned amplitude. The reaction-zone-conditioned heat-release fluctuation RMS is approximately 9.52 times the full-plane value, reflecting the dilution of localized heat release when averaged over the entire plane.

These results clarify the interpretation of the plane-averaged coupling measures used further in this work. The full-plane pressure signal represents the broader compressive response of the domain, whereas the full-plane heat-release signal represents an area-normalized net reacting source. Reaction-zone conditioning therefore strengthens the local heat-release amplitude while preserving its temporal ordering. Pressure remains more sensitive to spatial averaging because it is distributed beyond the chemically active layer.

\begin{figure*}[htbp]
  \centering
  \includegraphics[width=0.45\textwidth]{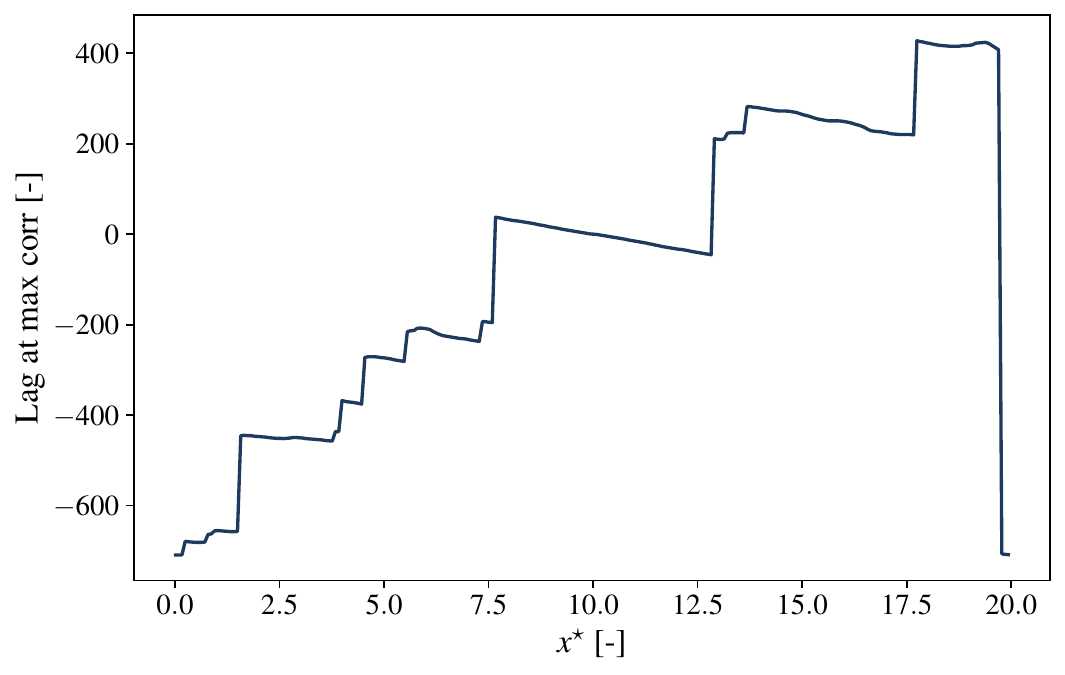}
  \includegraphics[width=0.45\textwidth]{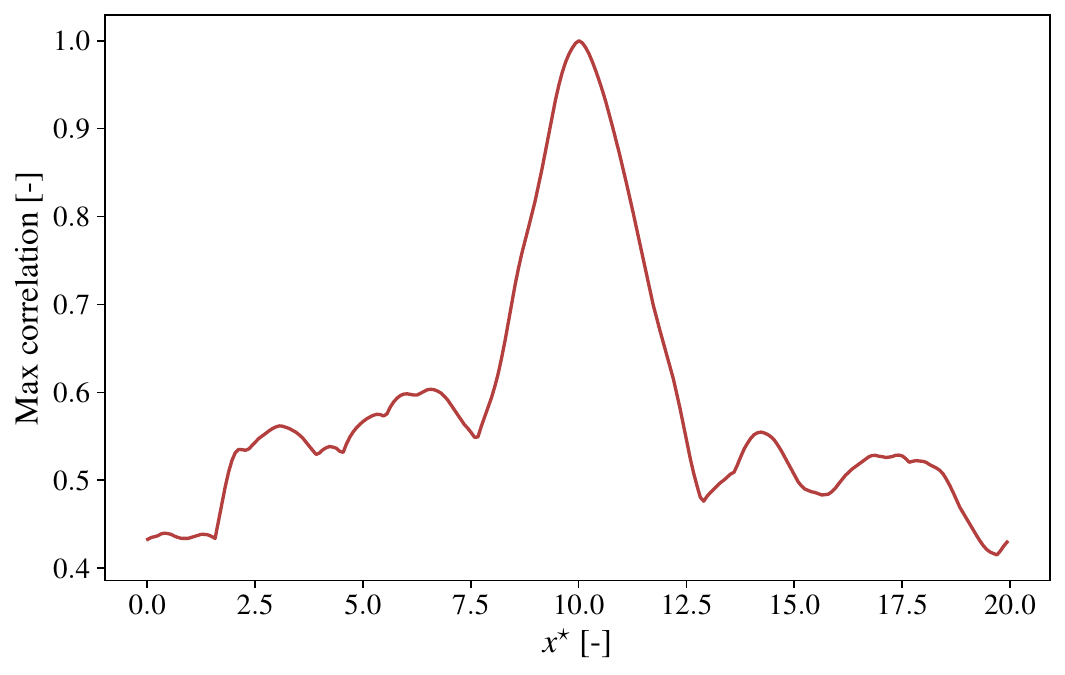}
  \caption{Near-field pressure-propagation measure based on two-point correlation of $y$-averaged pressure fluctuations. The left panel shows the lag at maximum correlation relative to a central reference location, and the right panel shows the corresponding maximum correlation.}\label{fig:pressure_propagation}
\end{figure*}

To further assess the streamwise organization of the pressure field, we compute a pressure-propagation measure from the $y$-averaged pressure field, $\langle p'(x,y,t)\rangle_y$. For a reference location $x_0$ near the center of the domain, the normalized two-point correlation is
\begin{equation}
  C_p(x,\tau)
  =
  \frac{
  \left\langle p_y'(x_0,t)\,p_y'(x,t+\tau)\right\rangle_t
  }{
  \sigma_{p_y(x_0)}\,\sigma_{p_y(x)}
  },
  \qquad
  p_y'(x,t)=\langle p'(x,y,t)\rangle_y,
\end{equation}
and we record the lag $\tau_{\max}(x)$ that maximizes $C_p$ along with the corresponding maximum correlation. The lag and correlation trends in \cref{fig:pressure_propagation} show finite correlation over the domain, with no clean isolated propagation branch away from the reference location. This supports the interpretation that the pressure signal is a mixed near-field compressive observable containing acoustic, shock-expansion, and hydrodynamic contributions. It is therefore interpreted as a pressure-response measure within the DNS plane rather than as a radiated acoustic-pressure field.

We can further project the time-series evolution into Fourier space to examine the spectral distribution of the source-side signals. Power spectral densities are computed from plane-averaged $p^\star$, $\dot{q}^\star$, $(\nabla\cdot\mathbf{u})^\star$, and $|\nabla Z|^{2\star}$, with cross-spectra and coherence assessing coupled oscillations. Frequencies are reported in Strouhal form $f^\star = \frac{f\,\delta_0}{U_c}$ to keep the nondimensionalization in this work consistent. Cumulative and band-integrated spectral energies summarize broadband redistribution.

For a discrete signal $x(t^\star)$, we define the Fourier transform $\widehat{x}(f^\star)$ and the power spectral density $S_{xx}(f^\star)=\langle |\widehat{x}(f^\star)|^2\rangle$. Cross-spectra are then defined as $S_{xy}(f^\star)=\langle \widehat{x}(f^\star)\widehat{y}^\ast(f^\star)\rangle$ and coherence defined as
\begin{equation}
  C_{xy}(f^\star)=\frac{|S_{xy}(f^\star)|^2}{S_{xx}(f^\star)S_{yy}(f^\star)}.
\end{equation}

\begin{figure*}[htbp]
  \centering
  \includegraphics[width=0.45\textwidth]{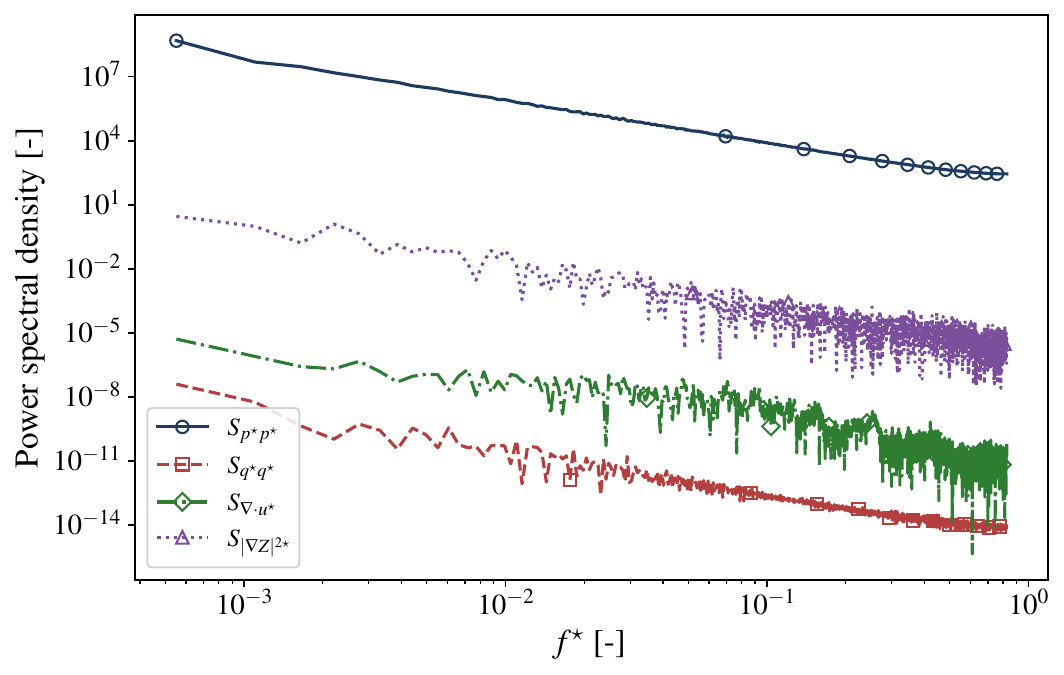}
  \includegraphics[width=0.45\textwidth]{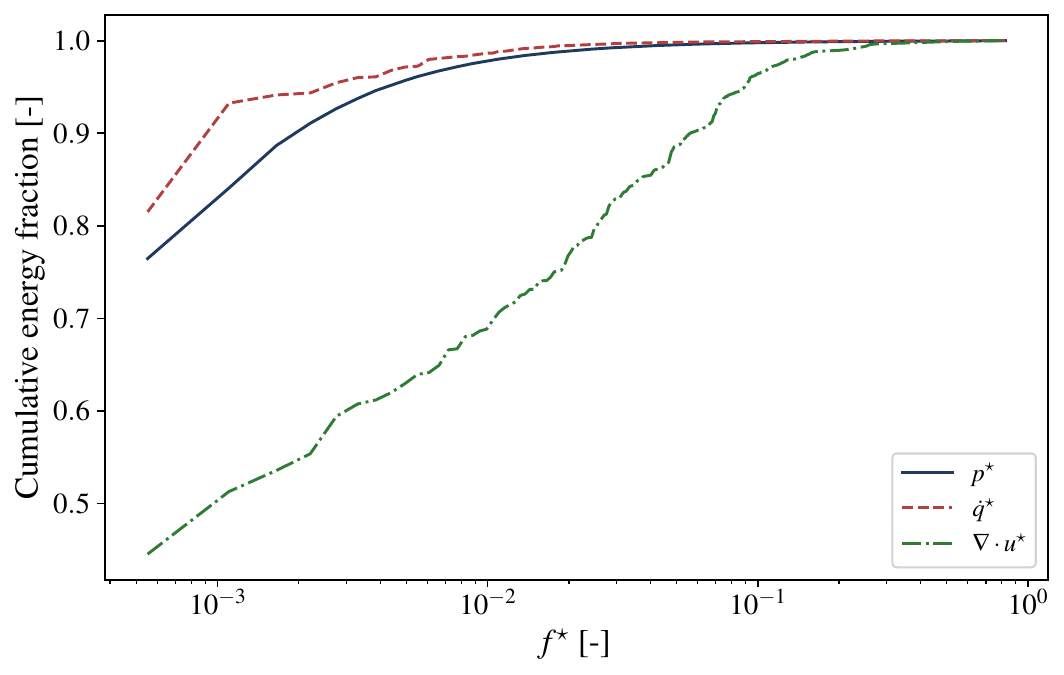}
  \\
  \includegraphics[width=0.45\textwidth]{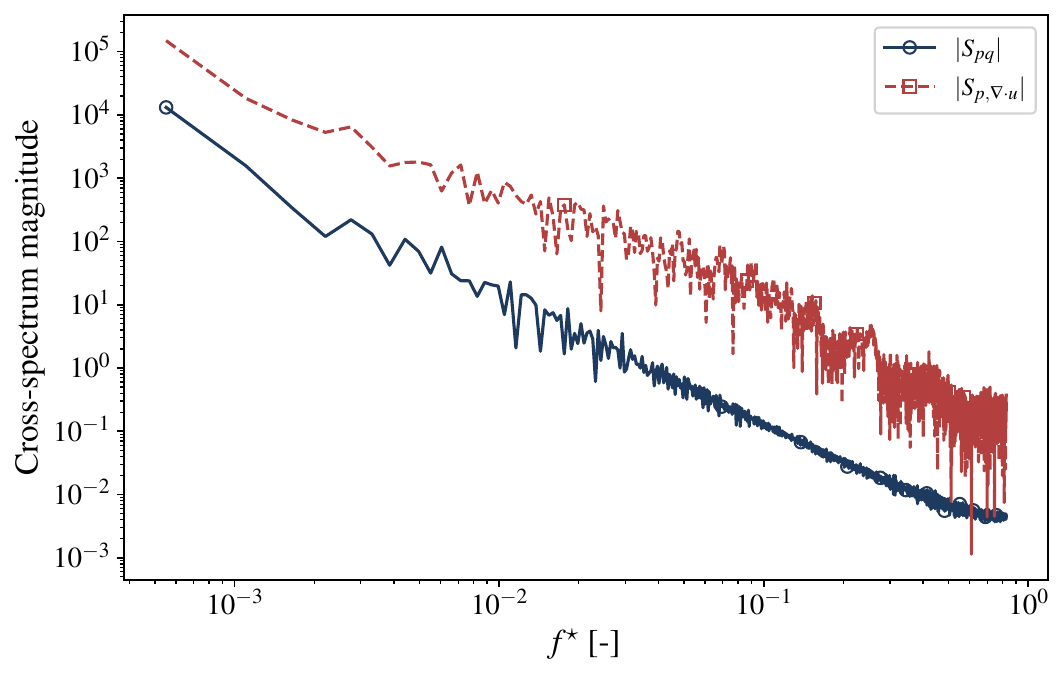}
  \includegraphics[width=0.45\textwidth]{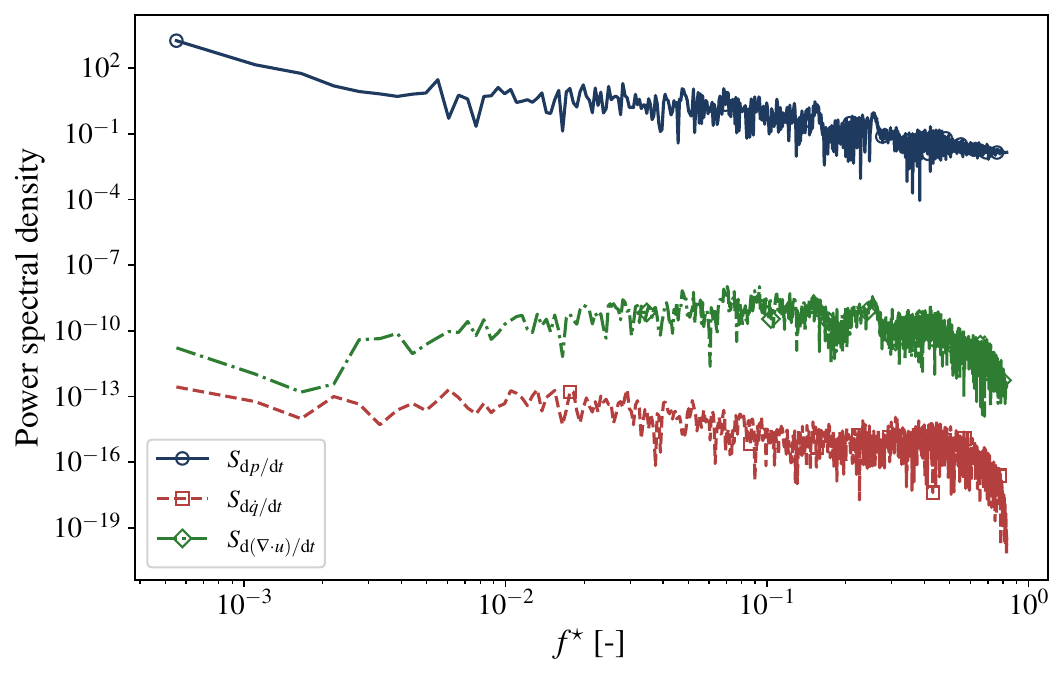}
  \caption{Spectra of plane-averaged signals and cumulative spectral energy (top), co-spectra (bottom left) and spectra of time derivatives (bottom right).}\label{fig:spectral_combo}
\end{figure*}

The smooth decay in the power spectra and cumulative energy (top row of \cref{fig:spectral_combo}) indicates that energy is distributed across a range of frequencies rather than concentrated in a few discrete peaks. The spectral density associated with pressure is the highest, consistent with the broader spatial support of pressure fluctuations seen in the mid-plane visualizations and with contributions from freestream and compressive motions. Dilatational PSD is also larger than that of heat release, supporting the corresponding mid-plane trends. For completeness, the spectral density of the scalar gradient is also shown, although a decomposition linking the scalar gradient with heat release and/or pressure is not performed in this work. The cross-spectra (bottom left panel of \cref{fig:spectral_combo}) and coherence (\cref{tab:coherence}) further indicate that pressure and heat release share only weakly phase-locked components, which is consistent with a temporally evolving shear layer in which heat-release bursts are not synchronized to a single dominant acoustic period. The PSDs of the corresponding unsteadiness in pressure, dilatation, and heat release follow the same ordering: pressure is highest, followed by dilatation, then heat release.

\begin{table}[htbp]
  \centering
  \caption{Coherence peak frequencies (nondimensional) for $p^\star$--$\dot{q}^\star$ and $p^\star$--$(\nabla\cdot\mathbf{u})^\star$.}\label{tab:coherence}
  \begin{tabular}{lcc}
    \hline
    Pair & Peak 1 & Peak 2 \\
    \hline
    $p^\star$--$\dot{q}^\star$ & 0.01928 & 0.03740 \\
    $p^\star$--$(\nabla\cdot\mathbf{u})^\star$ & 0.02709 & 0.01745 \\
    \hline
  \end{tabular}
\end{table}

The spectra show no narrow set of peaks in the plane-averaged observables, and the energy remains broadly distributed toward the lowest frequencies. This is consistent with a supersonic reacting shear layer in which compressibility and localized reaction events shape the near-field pressure response without producing a single large-scale periodic rollup signature. We also note that the cumulative energy fractions (top right of \cref{fig:spectral_combo}) are more uniformly distributed across frequencies for heat release and pressure, and slightly more biased toward the higher frequencies for dilatation, indicating that dilatation is comparatively weighted toward smaller temporal scales.

Time--frequency content is examined using a complementary pair of Short-time Fourier Transform (STFT) and Continuous Wavelet Transform (CWT) scalograms. The STFT uses a fixed window and therefore provides a clearer frequency-bin representation for comparison with the coherence peaks, while the CWT uses scale-dependent windows and is better suited for identifying transient, burst-like activity over multiple time scales \cite{Zhong2010,Farge1992,Mallat1999,Stankovic2013}. Thus, the STFT is used primarily to identify the frequency range over which energetic content persists, whereas the CWT is used to assess whether that content appears continuously or through intermittent localized events.

The STFT of $x$ with window $w$ is
\begin{equation}
X(t^\star,f^\star)=\sum_{n} x[n]\;w[n-t^\star]\;\exp(-2\pi i f^\star n),
\end{equation}
and the CWT uses a Morlet wavelet $\psi$,
\begin{equation}
W_x(a,b)=\int x(t^\star)\,\psi^\ast\!\left(\frac{t^\star-b}{a}\right)\,dt^\star,
\end{equation}
with scale $a$ mapped to frequency for scalogram visualization.

\begin{figure*}[htbp]
  \centering
  \includegraphics[width=0.49\textwidth]{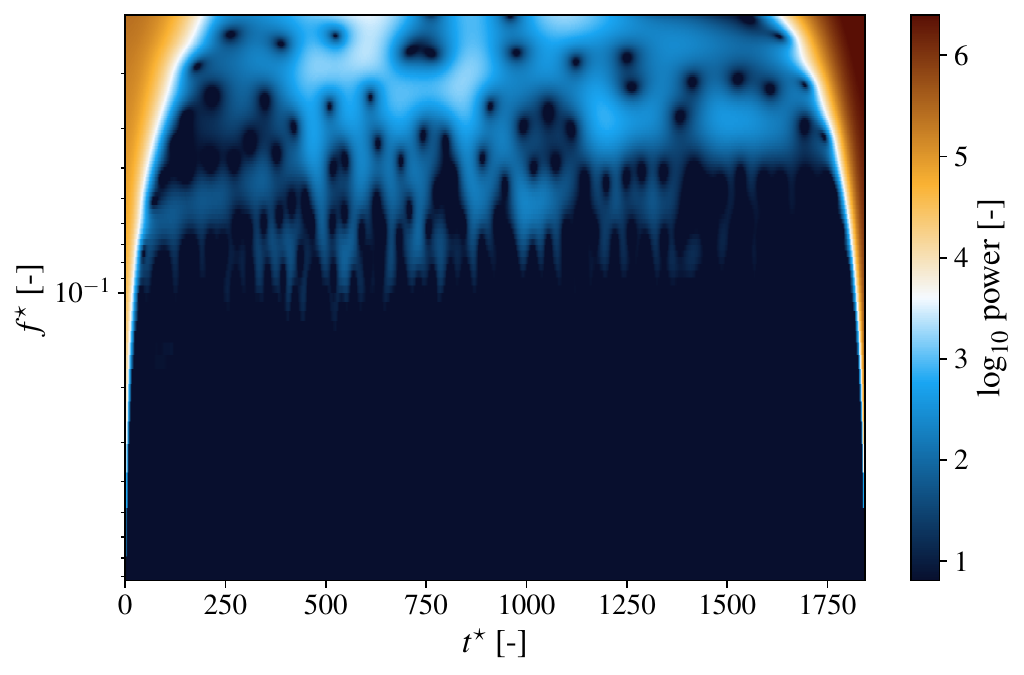}
  \includegraphics[width=0.49\textwidth]{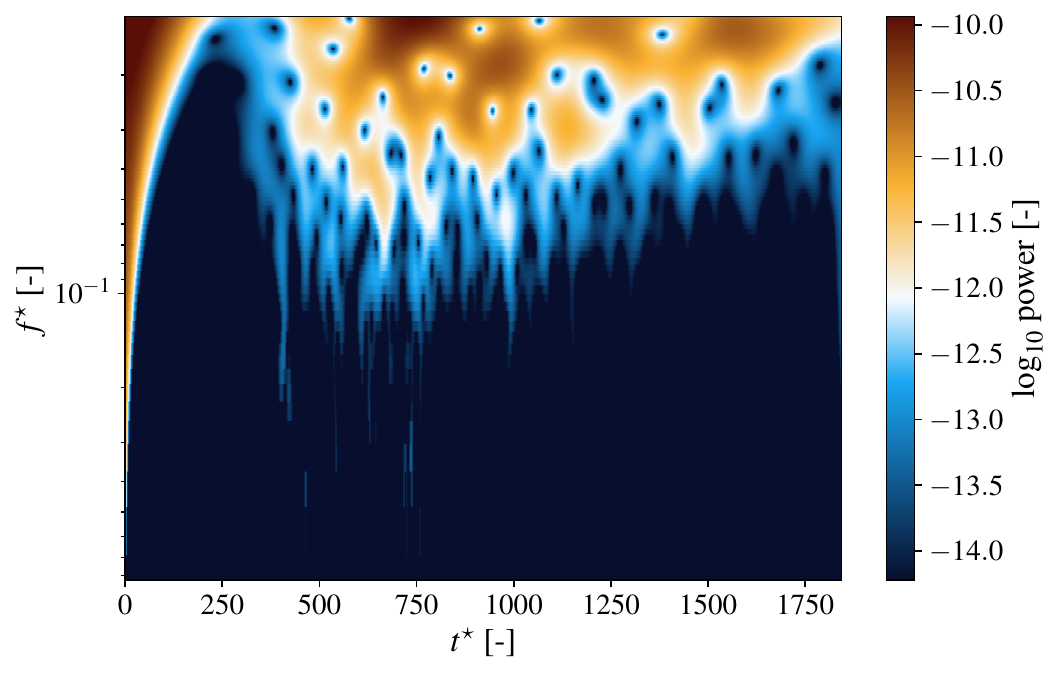}
  \\
  \includegraphics[width=0.49\textwidth]{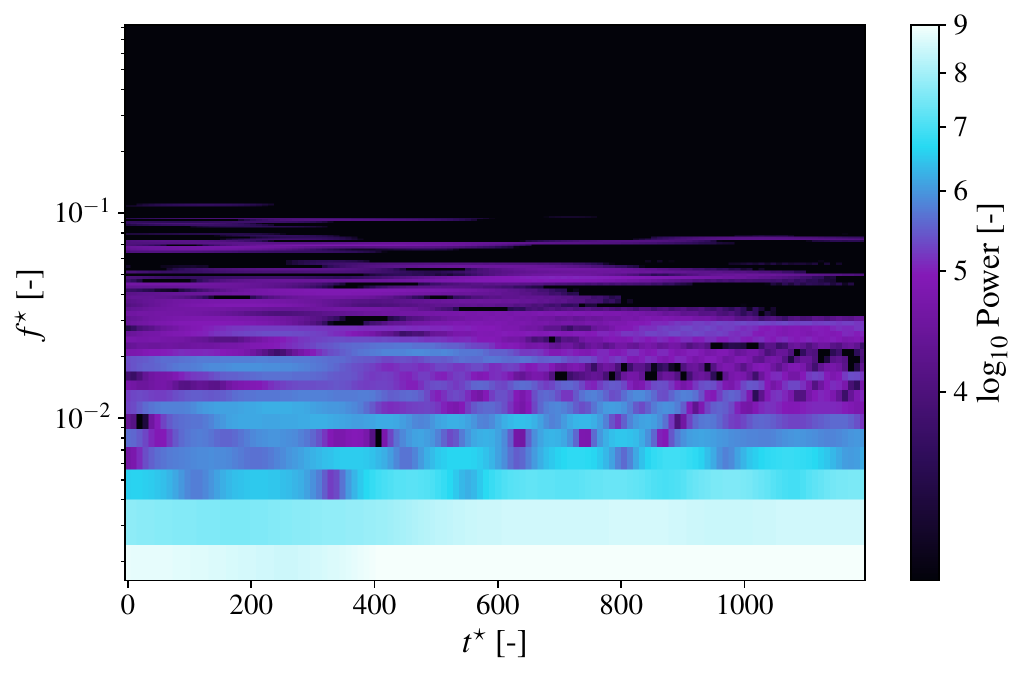}
  \includegraphics[width=0.49\textwidth]{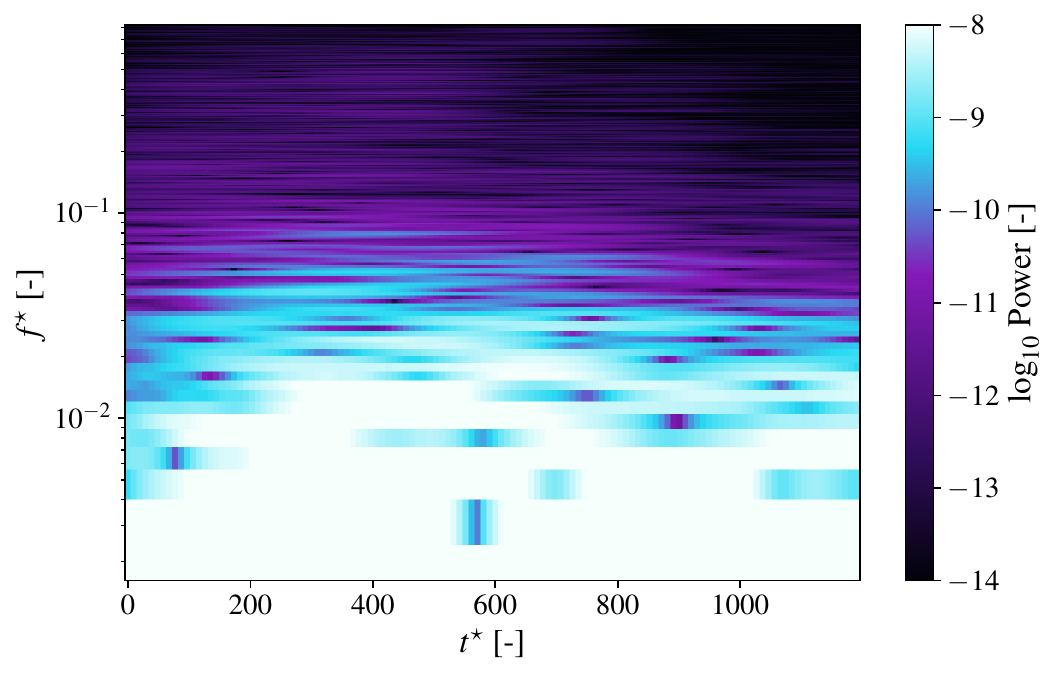}
  \caption{CWT scalograms (top) and STFT spectrograms (bottom) of $p^\star$ (left) and $\dot{q}^\star$ (right). The STFT emphasizes fixed-window frequency content, while the CWT emphasizes scale-localized temporal modulation.}\label{fig:timefreq}
\end{figure*}

The time--frequency maps in \cref{fig:timefreq} show that the strongest non-stationary activity remains concentrated at the low-frequency end of the spectrum. With the present STFT window, the frequency spacing is $\Delta f^\star \approx 6.5\times10^{-3}$. The pressure STFT contains sustained energy primarily below $f^\star \approx 0.026$, while the heat-release STFT extends over a somewhat broader low-frequency range, primarily below $f^\star \approx 0.058$. These ranges overlap the coherence peaks in \cref{tab:coherence}, which occur at $f^\star \approx 0.017$--$0.037$.

The CWT scalograms present the same dynamics from a scale-localized perspective. The pressure field shows low-frequency activity over much of the record, but the wavelet representation emphasizes intervals where this activity strengthens or weakens rather than displaying it as a uniformly persistent band. Heat release is more visibly intermittent, with localized bursts appearing over neighboring low-frequency scales. The apparent difference between the STFT and CWT therefore reflects the different time--frequency resolutions of the two transforms. The STFT emphasizes sustained frequency-bin energy within a fixed window, whereas the CWT emphasizes scale-localized temporal modulation. Together, the spectra, coherence, STFT, and CWT indicate weak pressure--heat release coupling associated with low-frequency broadband activity instead of a single sharply tuned mode.

The combination of spectra, co-spectra, and coherence further supports a broadband, weakly synchronized picture of the dynamics. Weak coherence implies that pressure and heat release fluctuate with variable phase, which is consistent with the absence of a stable feedback path in this boundary-free configuration.

To assess nonlinear phase coupling and effective near-field pressure response, we evaluate bicoherence, an impedance-like pressure--velocity ratio, and information-theoretic measures between pressure and heat-release signals. The bicoherence is computed as
\begin{equation}
b(f_1^\star,f_2^\star)=
\frac{
|\langle X(f_1^\star)X(f_2^\star)X^\ast(f_1^\star+f_2^\star)\rangle|^2
}{
\langle|X(f_1^\star)X(f_2^\star)|^2\rangle\,
\langle|X(f_1^\star+f_2^\star)|^2\rangle
}.
\end{equation}
Here, $X$ denotes the Fourier transform of the plane-averaged signal being analyzed, and $\langle\cdot\rangle$ denotes averaging over windowed time segments.

\begin{figure*}[htbp]
\centering
\includegraphics[width=0.45\textwidth]{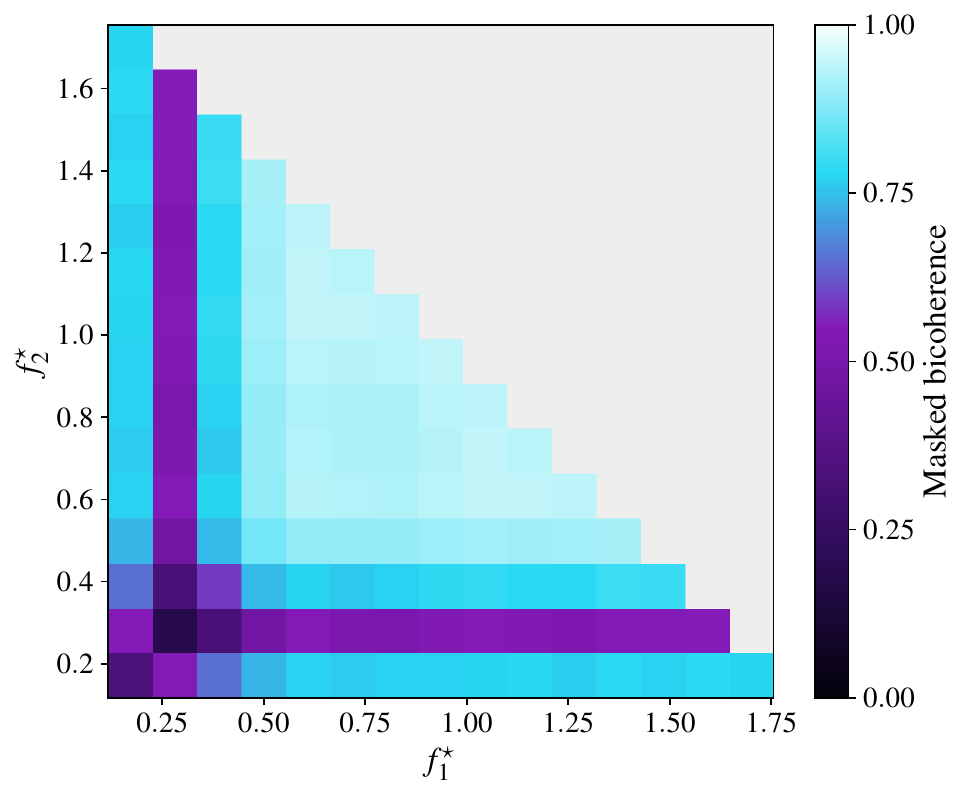}
\includegraphics[width=0.45\textwidth]{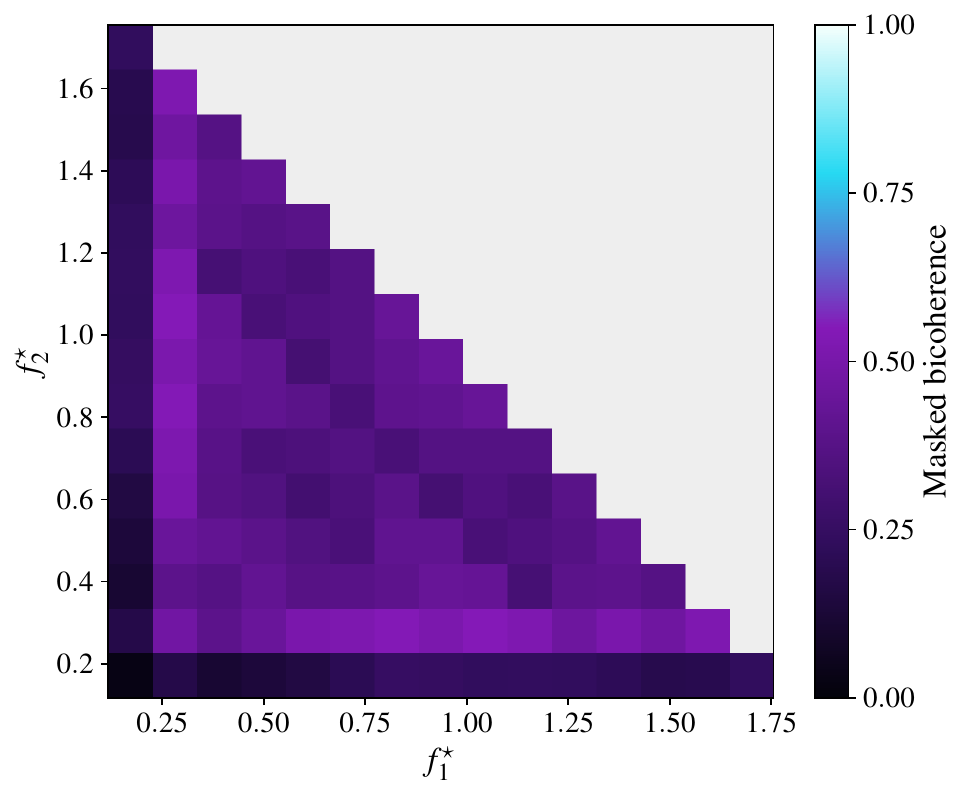}
\caption{Masked nonlinear bicoherence for plane-averaged $p^\star$ (left) and $\dot{q}^\star$ (right). Triads are shown only when all participating frequencies $f_1^\star$, $f_2^\star$, and $f_1^\star+f_2^\star$ exceed $3\times10^{-4}$ of the peak one-sided spectral power; gray regions indicate weak-power triads omitted from interpretation.}
\label{fig:bicoherence}
\end{figure*}

The masked bicoherence maps in \cref{fig:bicoherence} provide a frequency-plane representation of quadratic phase coupling in the plane-averaged pressure and heat-release signals. A nonzero value at $(f_1^\star,f_2^\star)$ indicates repeatable phase coupling among components satisfying $f_3^\star=f_1^\star+f_2^\star$. To avoid overinterpreting normalized bicoherence in weak-power regions, the maps retain only triads for which all participating frequencies exceed $3\times10^{-4}$ of the peak one-sided spectral power. Gray regions therefore denote weak-power triads that are omitted from interpretation.

Within the retained energetic low-frequency region, the pressure bicoherence is broadly elevated rather than concentrated around isolated triads. This indicates repeatable phase structure in the plane-averaged pressure signal. The heat-release bicoherence is lower in magnitude and more structured across the retained low-frequency region, suggesting weaker but more frequency-dependent nonlinear organization in the reacting source signal. Overall, the masked bicoherence supports broadband nonlinear phase organization at low frequencies, with no sharply localized harmonic cascade or single resonant triad in the retained energetic range.

\subsection{Impedance and Information-Theoretic Metrics}

To complement the spectral and coherence analysis, information-theoretic measures are computed to quantify statistical dependence and short-lag directional coupling between signals. The mutual information between two signals $x$ and $y$ is defined as
\begin{equation}
I(x;y)
= \sum_{i,j}
p_{xy}(i,j)
\ln\!\left(
\frac{p_{xy}(i,j)}{p_x(i)\,p_y(j)}
\right),
\end{equation}
where $p_{xy}$ denotes the joint probability distribution and $p_x$, $p_y$ are the corresponding marginals. This metric quantifies the total statistical dependence, including both linear and nonlinear contributions, between two signals. In the present study, $I(p;\dot{q})$ and $I(p;\nabla\cdot\mathbf{u})$ measure the overall coupling strength between pressure fluctuations and reaction-rate or compressibility dynamics, respectively.

Directional interactions are then assessed using transfer entropy at the minimum resolved temporal offset, corresponding to a one-sample lag $\Delta t^\star \approx 6.05\times10^{-1}$ under the fixed reference scaling used here. This choice avoids imposing an external acoustic, convective, or chemical time scale on the analysis and instead asks whether the immediately preceding state of one signal improves prediction of the next resolved state of the other. For two discrete signals, the transfer entropy from $y$ to $x$ is
\begin{equation}
T_{y\to x}
=
\sum
p(x_t,x_{t-1},y_{t-1})
\ln
\left(
\frac{p(x_t \mid x_{t-1},y_{t-1})}
     {p(x_t \mid x_{t-1})}
\right),
\end{equation}
which is equivalent to the conditional mutual information $T_{y\to x}=I(x_t; y_{t-1} \mid x_{t-1})$. Transfer entropy therefore quantifies how much knowledge of the past of $y$ improves prediction of the present state of $x$, beyond what is explained by the past of $x$ alone. Thus, $T_{\dot{q}\to p}$ measures whether past heat-release fluctuations improve prediction of present pressure fluctuations, while $T_{p\to \dot{q}}$ measures the reverse predictive direction.

The information-theoretic metrics show $I(p;\dot{q}) = 1.1368 > I(p;\nabla\cdot\mathbf{u}) = 0.3761$, indicating that pressure fluctuations share more overall statistical structure with heat release than with dilatation. The transfer entropy values also show directional asymmetry, with $T_{p\to \dot{q}} = 1.8297\times10^{-2}$ exceeding $T_{\dot{q}\to p} = 8.2063\times10^{-3}$, so past pressure fluctuations contain more predictive information about future heat release than vice versa. The one-sample value is interpreted as a short-lag predictive measure rather than as a unique physical delay. Lag-sensitivity checks over larger offsets produced changes in the absolute transfer-entropy magnitudes, as expected for a histogram-based estimator, but preserved the qualitative directional asymmetry between the two signals. The values are therefore used to summarize local predictive coupling between pressure and heat release, rather than to identify a causal propagation pathway.

In addition, a nondimensional frequency-dependent impedance-like pressure--velocity ratio is computed from the detrended plane-averaged pressure and streamwise velocity fluctuations as
  \begin{equation}
    Z^\star(f^\star)
    =
    \frac{\widehat{p^{\star\prime}}(f^\star)}
         {\widehat{u^{\star\prime}}(f^\star)},
  \end{equation}
where $\widehat{\cdot}$ denotes the Fourier transforms of the fluctuating pressure and velocity signals. In the present boundary-free configuration, this ratio characterizes the dynamic relationship between plane-averaged pressure and velocity. It is a near-field response measure for the sampled plane, distinct from boundary impedance or far-field radiation.

\begin{figure*}[htbp]
  \centering
  \includegraphics[width=0.75\textwidth]{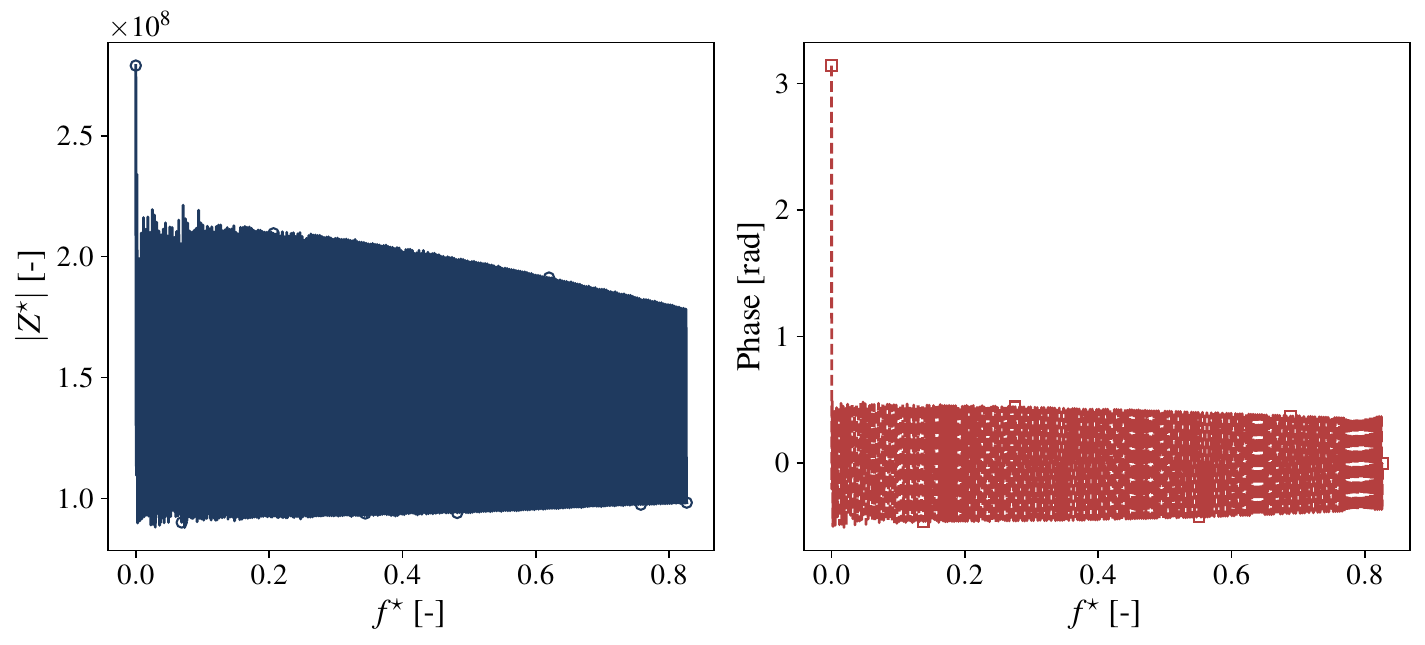}
  \caption{Impedance magnitude (left) and phase (right) for plane-averaged $p^\star$ and $u^\star$}\label{fig:impedance}
\end{figure*}

The magnitude $|Z^\star(f^\star)|$ in \cref{fig:impedance} decreases gradually with frequency and lacks sharp resonance peaks in the plane-averaged response. The phase varies within an approximately $[-\pi/6,\pi/6]$ window without large shifts of order $\pi$ that would suggest standing-wave behavior. These features are consistent with a broadband compressive response in the sampled plane.

Alongside the coherence, the impedance-like behavior indicates that combustion contributes to specific coupled frequency bands in the pressure field while the overall pressure--velocity dynamics remain broadband.

\subsection{Intermittency and burst statistics}
Burst events are defined from the plane-averaged heat-release activity signal. In the present implementation, this signal is the temporally mean-subtracted $\dot{q}_{\mathrm{rms}}^\star(t^\star)$, and burst intervals are contiguous times for which this signal exceeds its 70th percentile. Conditional PDFs of $p^\star$ and $(\nabla\cdot\mathbf{u})^\star$ are then computed separately over burst and non-burst samples, and are used to quantify intermittency effects and isolate compressive activity associated with strong heat-release events \cite{Onorato2000}. Sensitivity checks using thresholds between the 60th and 95th percentiles produced the same qualitative separation between burst-conditioned and non-burst pressure/dilatation statistics; the 70th percentile is therefore used as a representative threshold that retains sufficient samples while isolating elevated heat-release activity.

Formally, bursts are contiguous intervals $\mathcal{B}_k=[t_i^\star,t_j^\star]$ where $a_{\dot{q}}^\star(t^\star)>\theta$, with $a_{\dot{q}}^\star=\dot{q}_{\mathrm{rms}}^\star-\langle\dot{q}_{\mathrm{rms}}^\star\rangle_t$ and $\theta$ set by the 70th percentile of $a_{\dot{q}}^\star$. The burst duration is $\Delta t_k^\star=t_j^\star-t_i^\star$, amplitude is $\max_{\mathcal{B}_k}a_{\dot{q}}^\star$, and waiting time is the gap between consecutive intervals. Conditional PDFs are computed using samples restricted to the set $\bigcup_k \mathcal{B}_k$.

\begin{figure*}[htbp!]
  \centering
  \includegraphics[width=0.45\textwidth]{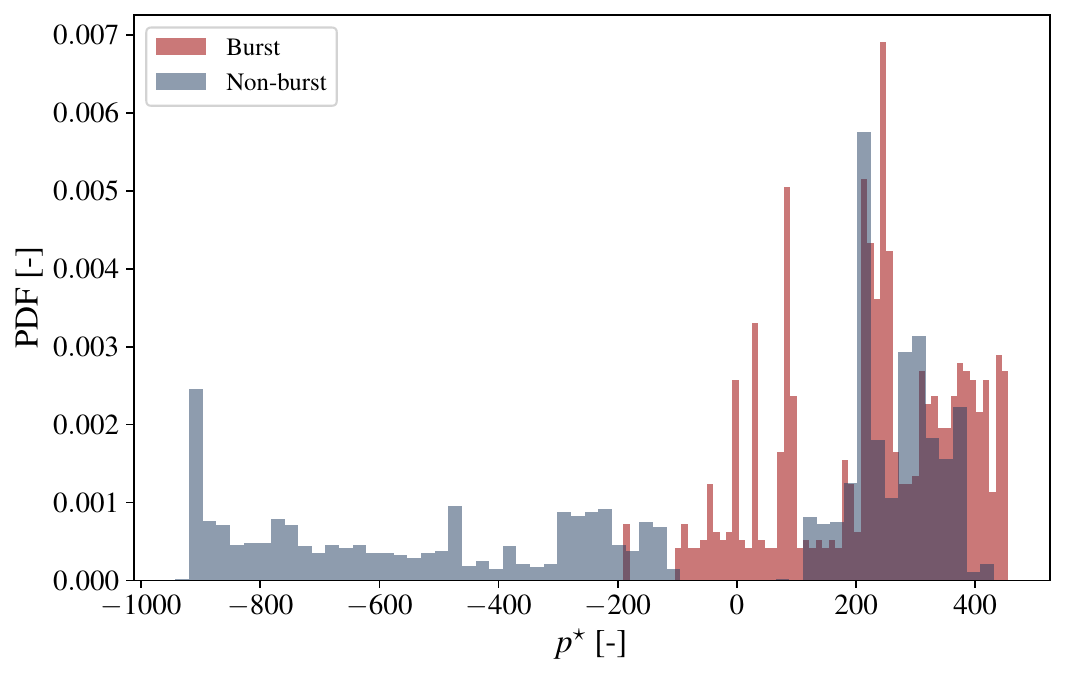}
  \includegraphics[width=0.45\textwidth]{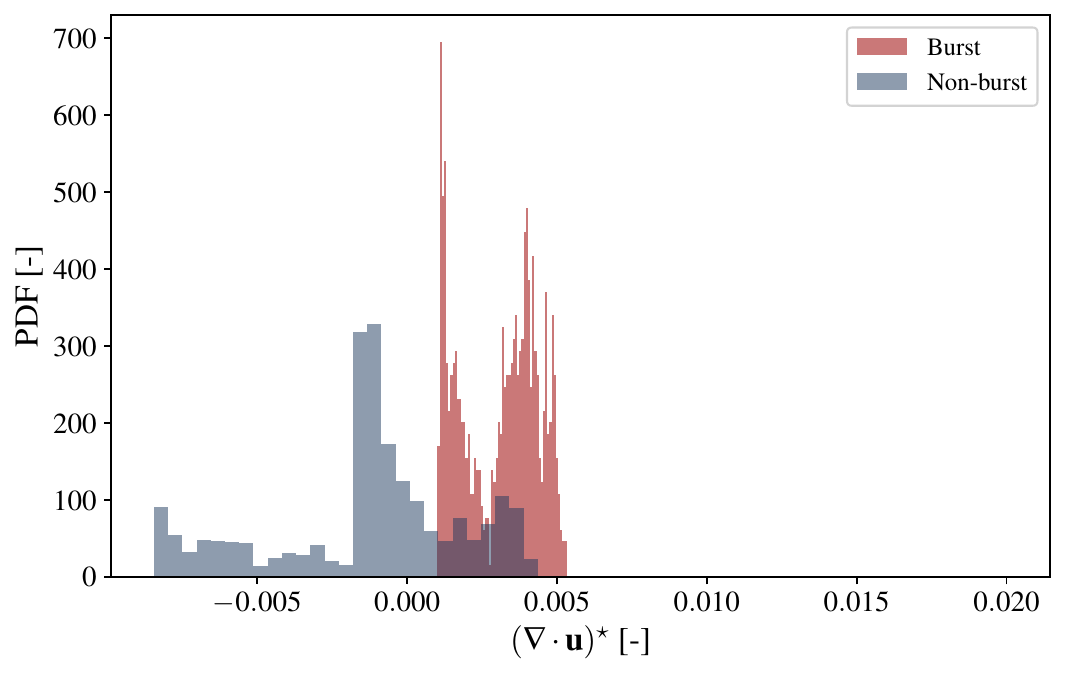}
  \caption{Conditional burst PDFs of $p^\star$ and $(\nabla\cdot\mathbf{u})^\star$}\label{fig:bursts_point}
\end{figure*}

The conditional burst PDFs in \cref{fig:bursts_point} show that pressure and dilatation excursions intensify during heat release bursts. The burst-conditioned dilatation distribution shifts toward positive values, indicating localized volumetric expansion consistent with rapid heat addition, while the pressure distribution is biased toward positive fluctuations relative to the non-burst state. These results demonstrate that combustion bursts are associated with statistically distinct flow states and connect the time--frequency signatures to the spatial source proxies: intermittent temporal bursts are associated with localized source patches within the shear layer.

Joint PDFs between pressure, heat release, and dilatation were also examined and were found to remain broad and diffuse, consistent with weak pointwise correlation and intermittent, band-limited coupling; they are not shown here in order to keep the presentation concise.

\subsection{Dynamical-systems characterization}
State-space descriptions used in thermoacoustic and combustion-noise analysis include phase portraits, delay embeddings, recurrence plots, Poincar\'e sections, and finite-time divergence proxies. Used together, these quantities distinguish weakly organized, burst-driven dynamics from low-dimensional periodic or resonant behavior.

\begin{figure*}[htbp]
  \centering
  \includegraphics[width=0.48\textwidth]{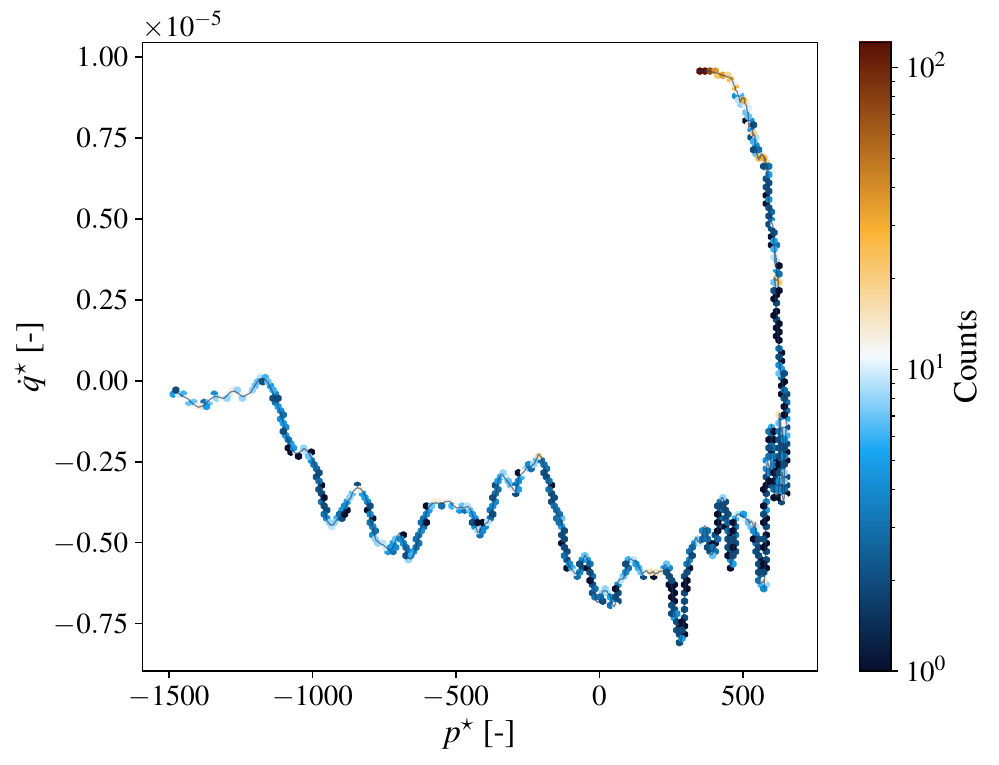}
  \includegraphics[width=0.48\textwidth]{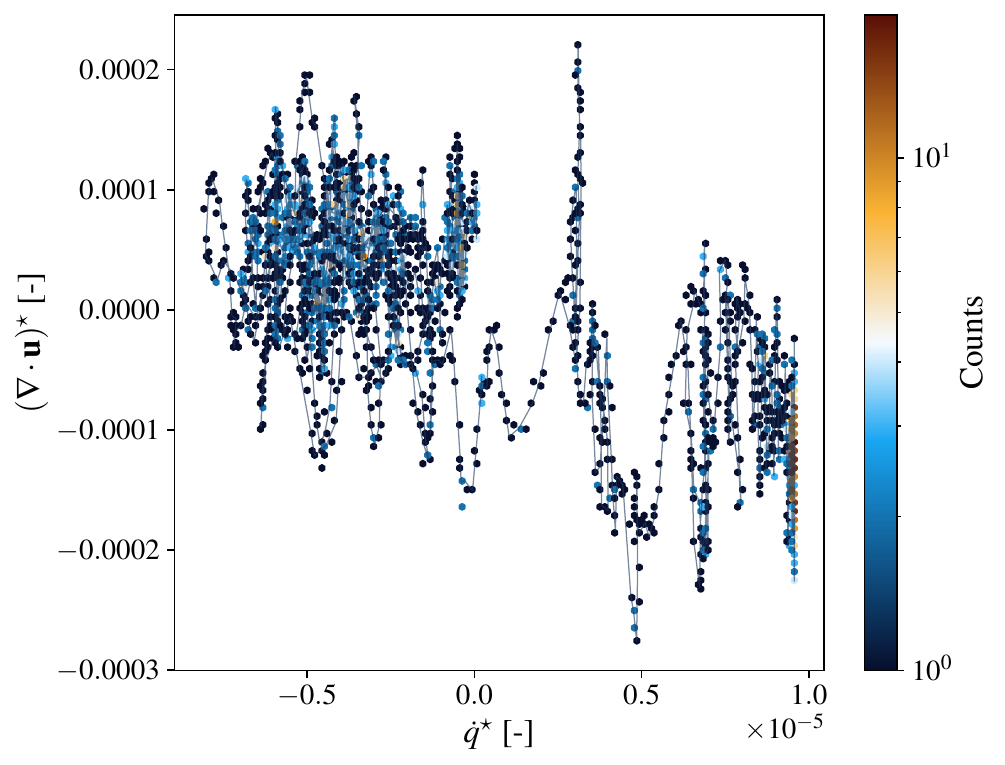}
  \caption{Phase portraits of $p^\star$--$\dot{q}^\star$ (left) and $\dot{q}^\star$--$(\nabla\cdot\mathbf{u})^\star$ (right)}\label{fig:phase_portraits}
\end{figure*}

Two-dimensional phase portraits are constructed to examine the instantaneous coupling between fluctuating variables. In \cref{fig:phase_portraits}, the $p^\star$--$\dot{q}^\star$ portrait forms a curved, asymmetric cloud rather than a narrow line or closed loop, indicating nonlinear and state-dependent interaction between pressure and heat release. Negative pressure excursions are predominantly associated with negative heat-release fluctuations, whereas positive pressure states span a broader range of combustion response.

The $\dot{q}^\star$--$(\nabla\cdot\mathbf{u})^\star$ portrait displays broader clustering with weaker geometric alignment, consistent with volumetric expansion accompanying heat release but with substantial variability due to additional hydrodynamic effects. Taken together, these phase-space structures suggest that combustion--pressure coupling is nonlinear and regime-dependent rather than globally linear. They are also qualitatively consistent with the information-theoretic result that pressure shares more overall statistical structure with heat release than with dilatation.

To assess whether heat-release fluctuations revisit similar dynamical states, a recurrence plot is constructed from the delay-embedded $\dot{q}^\star$ signal. For each time index $i$, the embedded state is
\begin{equation}
\mathbf{x}_i =
\left[\dot{q}^\star(t_i^\star),
\dot{q}^\star(t_i^\star+\tau^\star)\right],
\end{equation}
and the recurrence matrix is defined as
\begin{equation}
R_{i,j} =
\Theta\!\left(\varepsilon-\|\mathbf{x}_i-\mathbf{x}_j\|\right),
\end{equation}
where $\Theta(\cdot)$ is the Heaviside function and $\varepsilon$ is the recurrence threshold. A point in the recurrence plot therefore marks a pair of times at which the heat-release dynamics occupy nearby states in reconstructed phase space.

\begin{figure}[htbp]
  \centering
  \includegraphics[width=0.8\linewidth]{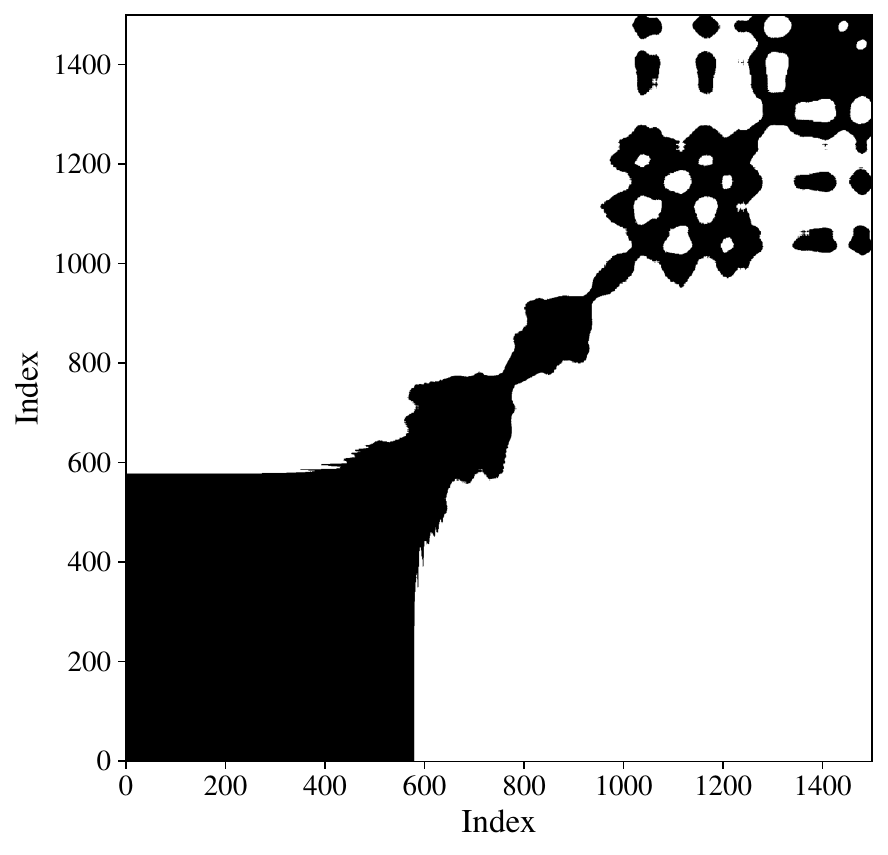}
  \caption{Recurrence plot of $\dot{q}^\star$}\label{fig:recurrence}
\end{figure}

The main diagonal corresponds to the trivial recurrence of each state with itself. Off-diagonal points indicate nontrivial returns, where the heat-release signal revisits a similar state at a later or earlier time. Short diagonal segments show intervals over which two portions of the signal evolve similarly for a finite time, while isolated points or compact clusters indicate brief returns that do not persist. In \cref{fig:recurrence}, the recurrence structure is therefore intermittent rather than periodic: the plot contains localized clusters and short diagonal features, but it does not show long, regularly spaced diagonal bands. This indicates that the heat-release dynamics repeatedly organize around similar burst states, but those returns are irregular and short-lived rather than locked to a stable cycle. Thus, the recurrence plot in the reacting layer reveals burst-driven temporal organization, with episodic revisitation of similar heat-release states, but it does not behave like a low-dimensional periodic oscillator or a feedback-sustained thermoacoustic mode.

\begin{figure*}[htbp]
  \centering
  \includegraphics[width=0.99\textwidth]{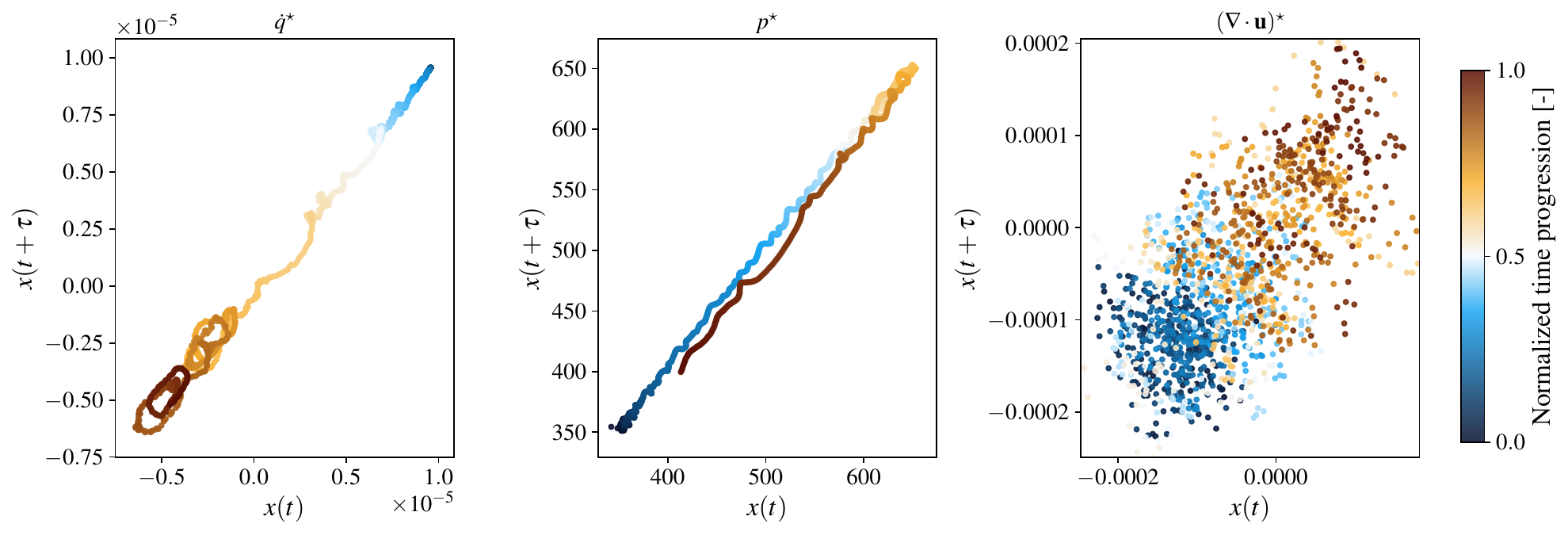}
  \caption{Delay embeddings for heat release rate (left), pressure (middle), and dilatation (right), colored by normalized time progression.}\label{fig:delay_embeddings}
\end{figure*}

To examine the intrinsic temporal structure of the signals, two-dimensional delay embeddings were constructed using
\begin{equation}
  \mathbf{x}(t^\star)=[x(t^\star),x(t^\star+\tau^\star)],
\end{equation}
where $\tau^\star$ is a fixed time delay. This representation provides a geometric view of the system's dynamics in reconstructed phase space. In \cref{fig:delay_embeddings}, the heat-release embedding exhibits clustered, loop-like structures, suggesting intermittent revisitation of similar states associated with burst dynamics. The pressure embedding follows a narrower, elongated trajectory, consistent with smoother temporal evolution. In contrast, the dilatation embedding is more diffuse, with a broader spread in phase space, suggesting greater variability in compressibility-driven fluctuations.

The embeddings therefore reinforce the broader picture that heat release is intermittent but organized, pressure is comparatively smoother and more constrained, and dilatation remains more broadly distributed in state space. None of the embeddings exhibits the compact closed geometry expected of a low-dimensional resonant oscillator.

\begin{figure*}[htbp]
  \centering
  \includegraphics[width=0.45\textwidth]{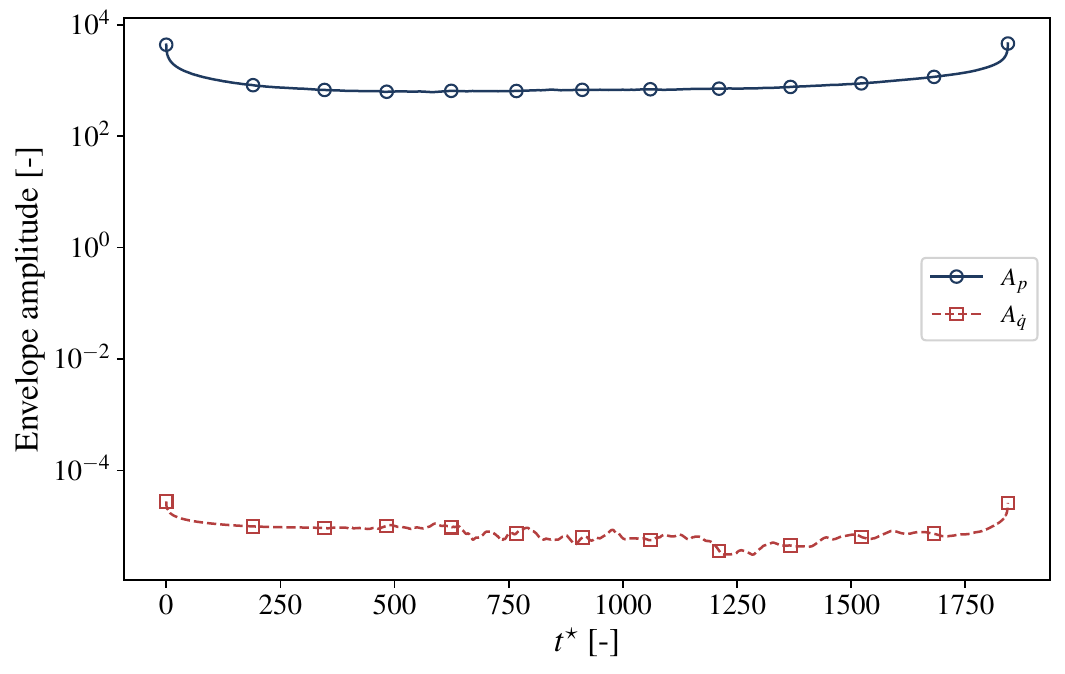}
  \includegraphics[width=0.45\textwidth]{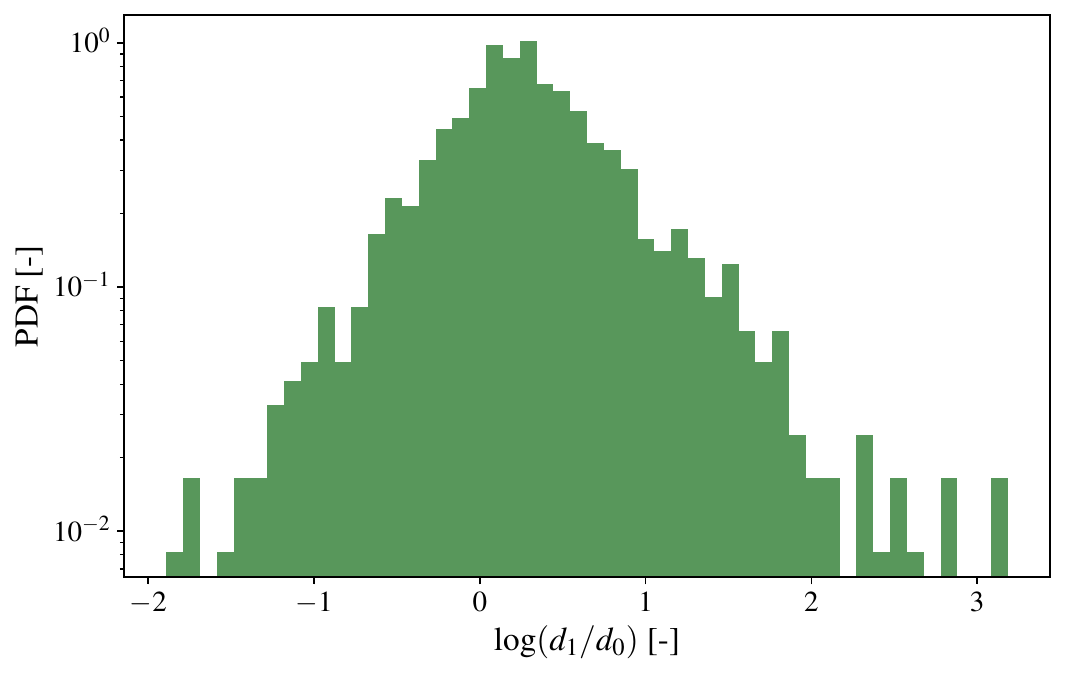}
  \caption{Hilbert envelopes and finite-time divergence proxy}\label{fig:hilbert_divergence}
\end{figure*}

Hilbert transforms are used here to construct analytic signals and extract envelope amplitudes. For a real-valued signal $x^\star(t^\star)$, the analytic signal is
\begin{equation}
  z^\star(t^\star)
  =
  x^\star(t^\star)
  +
  i\,\mathcal{H}\{x^\star(t^\star)\},
\end{equation}
and the instantaneous envelope amplitude is
\begin{equation}
  A_{x^\star}(t^\star)
  =
  |z^\star(t^\star)|
  =
  \sqrt{
  x^{\star 2}(t^\star)
  +
  \mathcal{H}\{x^\star(t^\star)\}^2 }.
\end{equation}
The envelopes $A_{p^\star}(t^\star)$ and $A_{q^\star}(t^\star)$ quantify amplitude modulation independent of carrier oscillations. Temporal coincidence of envelope spikes indicates amplification of pressure fluctuations concurrent with nondimensional heat-release activity, whereas pressure-envelope excursions without corresponding heat-release amplification suggest additional hydrodynamic contributions. The envelopes in \cref{fig:hilbert_divergence} show broad amplitude modulation in both $q^\star$ and $p^\star$, with pressure exhibiting wider intervals of elevated envelope response than the more intermittent heat-release signal.

To assess short-horizon sensitivity of the combustion dynamics, a finite-time divergence metric is computed from the delay-embedded heat-release signal $q^\star(t^\star)$ as
\begin{equation}
  \mathbf{x}(t^\star) =
  \big(q^\star(t^\star),\,
       q^\star(t^\star+\tau^\star),\,
       q^\star(t^\star+2\tau^\star)\big),
\end{equation}
where $\tau^\star$ denotes the nondimensional delay. For pairs of nearby embedded states with initial separation $d_0 = \|\mathbf{x}_i - \mathbf{x}_j\|$, the separation after a finite evolution interval is measured as $d_1 = \|\mathbf{x}_{i+\Delta} - \mathbf{x}_{j+\Delta}\|$, and the divergence is defined as $\Delta = \log\!\left(d_1/d_0\right)$. The resulting probability density of $\Delta$ characterizes short-time contraction ($\Delta < 0$) and separation ($\Delta > 0$) in the reconstructed phase space of $q^\star$. Positive support indicates intermittent sensitivity of the nondimensional heat-release dynamics, consistent with burst-driven temporal variability.

The finite-time divergence distribution is centered slightly above zero with finite positive and negative support, indicating intermittent short-horizon trajectory separation in the delay-embedded heat-release dynamics without evidence of sustained exponential growth. In combination with the Hilbert envelopes, this is consistent with the idea that strong heat-release bursts coincide with temporarily elevated dynamical sensitivity, while the overall bounded distribution remains compatible with recurrent, statistically stable evolution.

\begin{figure*}[htbp]
  \centering
  \includegraphics[width=0.45\textwidth]{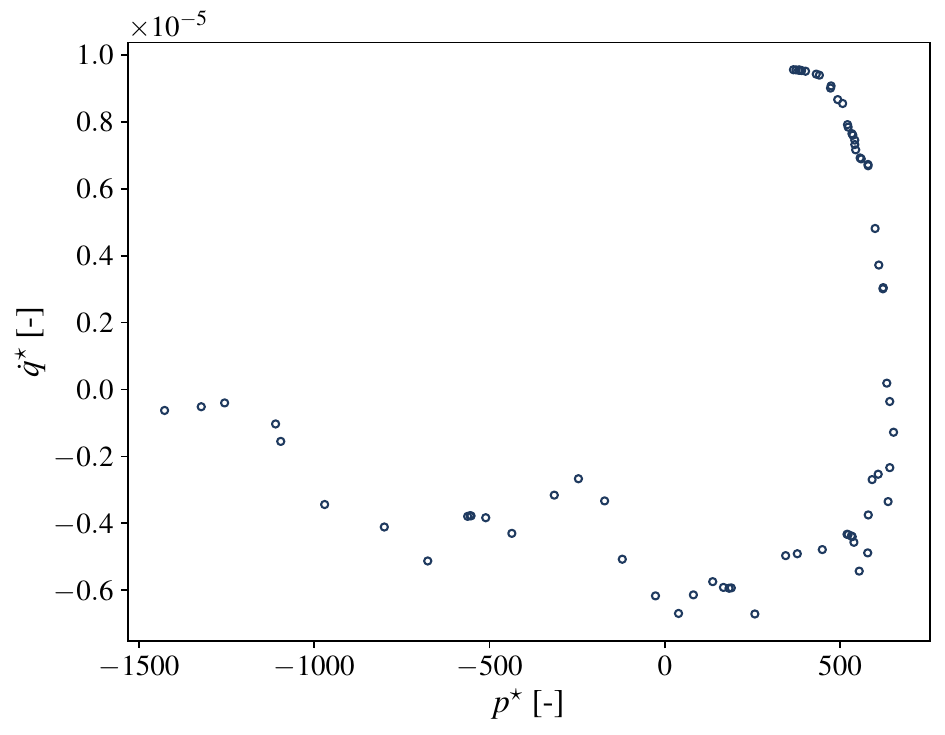}
  \includegraphics[width=0.45\textwidth]{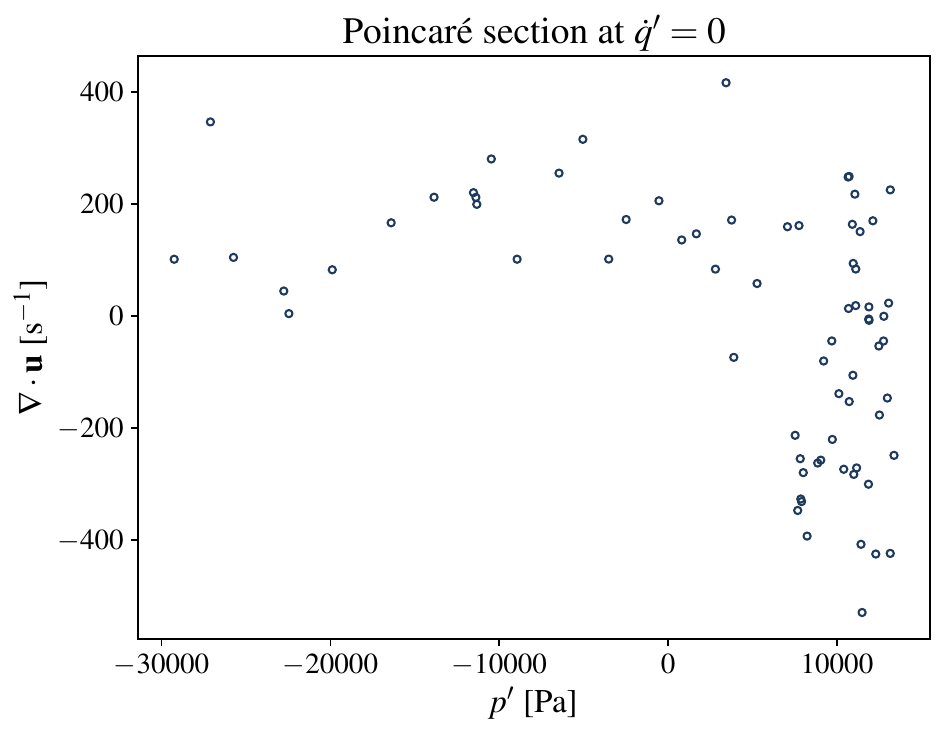}
  \caption{Poincar\'e sections for $p^\star$--$\dot{q}^\star$ (left) and $p^\star$--$(\nabla\cdot\mathbf{u})^\star$ (right).}\label{fig:poincare}
\end{figure*}

Poincar\'e sections in \cref{fig:poincare} likewise show diffuse crossings rather than compact repeated intersections, which is again consistent with weakly organized, broadband dynamics instead of a low-dimensional resonant cycle.

\section{Conclusions}
This study examined near-field combustion-noise source dynamics in a supersonic reacting shear layer. The source-side picture is controlled by the spatial separation between localized thermochemical activity and a broader compressive pressure field. Heat release and scalar-gradient activity remain concentrated within the reacting shear layer, while pressure and dilatation occupy a larger portion of the sampled plane because they include shock-expansion and freestream compressibility contributions.

The temporal analysis shows that pressure, heat release, and dilatation are organized primarily through broadband low-frequency activity. Coherence and time--frequency measures identify selected nondimensional frequency bands where heat release and pressure are statistically coupled, but the coupling is weak and intermittent. Conditional Rayleigh index analysis further shows that positive heat release--pressure phasing intensifies during high-activity intervals, connecting the low-frequency pressure response to burst-like combustion events. The source-radiation-potential projection indicates a moderate low-frequency transverse bias of the heat-release source distribution, while the impedance-like pressure--velocity ratio has a smooth broadband magnitude and phase response in the plane-averaged signal. The pressure-propagation measure indicates a mixed near-field compressive observable rather than an isolated radiated acoustic branch.

The nonlinear and dynamical measures support the same physical interpretation. Masked bicoherence identifies broadband low-frequency phase organization in the retained energetic triads, while recurrence, delay embeddings, finite-time divergence, and Hilbert envelopes show intermittent amplitude modulation and transient short-time sensitivity. Overall, combustion intermittency organizes near-field pressure fluctuations through localized source patches, broad low-frequency bands, and transient burst events. The conclusions are source-side conclusions for the sampled DNS plane and provide a foundation for future observer-dependent radiation calculations.

\section*{Acknowledgments}

The computing power for this study was provided by the Phoenix Computing Cluster as a part of Georgia Tech's Partnership for Advanced Computing Environment, and is gratefully acknowledged. The authors would like to thank Dr. Dhruv Purushotham at the High Performance Computing Lab at Georgia Tech for proofreading and critiquing this work.

\section*{Author Declarations}

\subsection*{Author Contributions}
Sriram P. Kalathoor: Conceptualization, Methodology, Software, Formal analysis, Investigation, Visualization, Writing -- original draft. Joseph C. Oefelein: Conceptualization, Resources, Supervision, Writing -- review and editing.

\subsection*{Funding}
This research received no specific grant from any funding agency, commercial or not-for-profit sectors.

\subsection*{Conflict of Interest}
The authors have no conflicts to disclose.

\section*{Data Availability}
The data that support the findings of this study are available from the corresponding author upon reasonable request.

\bibliographystyle{unsrtnat}
\bibliography{refs}

\end{document}